\documentclass[a4paper,12pt]{article}
\pdfoutput=1
\usepackage{hyperref}
\usepackage{epsfig}
\usepackage{amssymb}
\usepackage{color}
\usepackage{setspace}

\usepackage{amsmath}
\usepackage{amssymb}
\usepackage{graphicx}
\usepackage{geometry}
\usepackage{subfigure}
\usepackage{url}
\usepackage{slashed}

\newcommand{\be}{\begin{equation}}
\newcommand{\ee}{\end{equation}}
\newcommand{\br}{\begin{eqnarray}}
\newcommand{\bea}{\begin{eqnarray}}
\newcommand{\eea}{\end{eqnarray}}
\newcommand{\er}{\end{eqnarray}}
\newcommand{\ba}{\begin{array}}
\newcommand{\ea}{\end{array}}
\newcommand{\bi}{\begin{itemize}}
\newcommand{\ei}{\end{itemize}}
\newcommand{\bn}{\begin{enumerate}}
\newcommand{\en}{\end{enumerate}}
\newcommand{\bc}{\begin{center}}
\newcommand{\ec}{\end{center}}

\newcommand{\beq}{\begin{equation}}
\newcommand{\eeq}{\end{equation}}

\newcommand{\ET}{{\hspace{-0.25cm} E_{T}}}

\newcommand{\U}{\scriptscriptstyle U}
\newcommand{\D}{\scriptscriptstyle D}
\newcommand{\Q}{\scriptscriptstyle Q}
\newcommand{\EE}{\scriptscriptstyle E}
\newcommand{\E}{\scriptscriptstyle E}

\newcommand{\gsim}{\lower1.0ex\hbox{$\;\stackrel{\textstyle>}{\sim}\;$}}
\newcommand{\lsim}{\lower1.0ex\hbox{$\;\stackrel{\textstyle<}{\sim}\;$}}

\newcommand{\bs}{\begin{small}}
\newcommand{\es}{\end{small}}

\newcommand{\qui}{q_{{\scriptscriptstyle U}_{\!i}}}
\newcommand{\qdi}{q_{{\scriptscriptstyle D}_{\!i}}}

\def\mysection#1{{\bf #1.} }

\begin{document}
\pagestyle{empty}
\begin{center}
{\LARGE {\bf 
Higgs-boson production in association 
\\
\vspace{0.5cm}
with  a  Dark Photon in $e^ +e^ -$ collisions
}} \\
\vspace*{1.5cm}
{
 {\bf Sanjoy Biswas$^{{a}}$},
 {\bf Emidio Gabrielli$^{{b,c}}$},  {\bf Matti Heikinheimo$^{{c}}$}, 
{\bf and Barbara Mele$^{d}$}}\\

\vspace{0.5cm}
{\it
(a) KIAS,
85 Hoegi-ro, Dongdaemun-gu, 
Seoul 130-722, Republic of Korea}   \\[1mm]
{\it
 (b) Dipart. di Fisica Teorica, Universit\`a di 
Trieste, Strada Costiera 11, I-34151 Trieste, Italy and 
INFN, Sezione di Trieste, Via Valerio 2, I-34127 Trieste, Italy}  
\\[1mm]
{\it
 (c) NICPB, Ravala 10, Tallinn 10143, Estonia}  \\[1mm]
 {\it
 (d) INFN, Sezione di Roma, c/o Dipart. di Fisica, Universit\`a di Roma ``La Sapienza", \\ P.le Aldo Moro 2, I-00185 Rome, Italy} 

\vspace*{2cm}{\bf ABSTRACT}
\end{center}

\vspace{0.3cm}

We study  the production of a Higgs boson recoiling from  a massless invisible system
in  $e^ + e^ -$ collisions. This is a quite distinctive signature that can arise when the 
  Higgs boson is produced in association 
  with a  massless dark photon, which can happen  
 in BSM scenarios foreseeing an extra unbroken  $U(1)$ gauge group. Dark photons  can indeed  acquire  effective couplings to the Higgs boson as occurs in models recently proposed to generate exponentially-spread Yukawa couplings.
We analyze the signal  and corresponding backgrounds for $H\to b\bar{b}$, and estimate
ILC and FCC-ee sensitivities  in a model-independent way.

\vspace*{5mm}

\noindent

\vfill\eject

\pagestyle{plain}


\section{Introduction}
The LHC discovery  of the Higgs-boson resonance at 125 GeV \cite{Aad:2012tfa} has definitely strengthened our confidence in the Higgs mechanism as the origin of the electroweak (EW) symmetry breaking (EWSB) and fermion mass generation  \cite{Englert:1964et}.  All present data are well consistent with the Standard Model (SM)  expectations for the Higgs boson  \cite{ATLAS:h}-\cite{CMS:2014ega}, although there is still room for potential New Physics (NP) effects.

At the same time, 
the  absence  of any NP signal at the LHC Run I is 
causing considerable concern about the applicability of the naturalness criteria. 
The latter would require new phenomena at the TeV scale to stabilize the SM Higgs scalar potential against potentially large radiative corrections coming from NP energy thresholds.
The SM is also facing  the Flavor problem, 
which is related to the unexplained huge hierarchy in the fermion mass spectrum or, analogously, in the  Higgs Yukawa couplings. 

On another front, there is increasing evidence from astrophysical and cosmological observations of the existence of  Dark Matter (DM) in the Universe \cite{Ade:2013zuv,Planck:2015xua}, which is not  predicted in the SM. Independently of naturalness criteria, there might then be  NP above the EW scale that explains DM and, in some cases, could be tested at the LHC.

A common origin for DM and Flavor is also conceivable. One can postulate the existence of a hidden (dark) sector, where all these issues are addressed, which is composed of new fields that are SM singlets. The Higgs boson can then act as a portal to the dark sector \cite{Patt:2006fw}.
The Flavor and EWSB structures are indeed restricted to the 
Higgs couplings and mass, and are not related to  other SM couplings. On the other hand,  NP could well affect  the  Higgs-boson characteristics by smaller amounts  than the present LHC sensitivity in Higgs-boson data.

In this connection, in \cite{Gabrielli:2013jka} 
a new  paradigm has been proposed to generate exponentially-spread Yukawa couplings from gauge quantum numbers  in the dark sector. In this class of models, the  Flavor and chiral symmetry breaking (ChSB) take place in a dark sector, and are transmitted by radiative corrections to the observable sector through Higgs-portal type of interactions.
The Yukawa couplings  arise radiatively as effective couplings at low energy.
 The hidden sector consists of (stable) massive dark fermions 
(that are SM singlets and potential DM candidates), and a massless dark photon, the gauge boson of  an unbroken $U(1)_F$ gauge group in the dark sector. 
Chiral symmetry is spontaneously broken, and dark fermions obtain non-vanishing 
(flavor-dependent) masses via a non-perturbative mechanism involving $U(1)_F$ gauge interactions.
The resulting  chiral-symmetry and flavor breaking in the dark sector is then transferred to the  Yukawa-coupling sector at one-loop  via  scalar messenger fields that are  charged under both SM and $U(1)_F$ gauge interactions.
A similar framework has also been  explored in \cite{Ma:2014rua,Fraser:2014ija}, although  no unbroken $U(1)$ gauge sector is introduced in that case.

 The new unbroken $U(1)$ gauge group (and the corresponding massless dark photon) 
  is a crucial dynamical component of the model in  \cite{Gabrielli:2013jka}, but is  also  a common feature of various theories of new physics, including models with 
gauge-symmetry breaking of compact gauge groups, string-theory motivated phenomenological models, and models of interacting dark matter \cite{Holdom},\cite{Holdom:1985ag}. It is indeed conceivable that a  hidden sector  contains an extra long-range force.
Remarkably, being massless, an {\it on-shell} dark photon can be fully decoupled   from the SM quark and lepton sector at any order in perturbation theory  \cite{Holdom}.
This is not true for a massive dark-photon, due to a potential 
tree-level mixing with the photon field.
Most of present astrophysical and accelerator constraints  \cite{Abazov:2009hn}  apply to {\it massive} dark-photon couplings, and can be evaded in a massless dark-photon scenario. This allows for potentially large dark-photon couplings  to the dark sector, that might  also lead to observable new signatures  at colliders \cite{Gabrielli:2013jka}.

\begin{figure}
\begin{center}
\hskip -0.5cm
\includegraphics[width=0.27\textwidth]{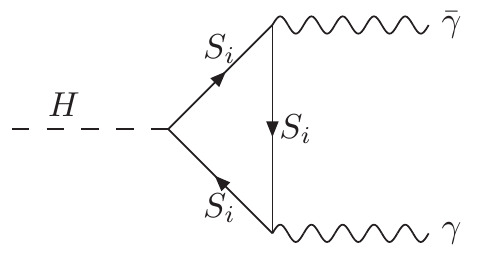}
\includegraphics[width=0.27\textwidth]{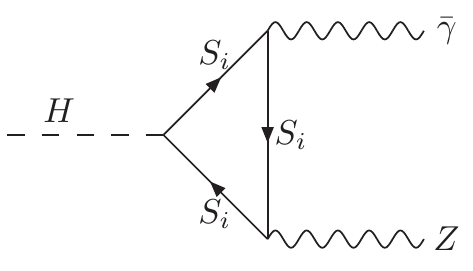} 
\includegraphics[width=0.27\textwidth]{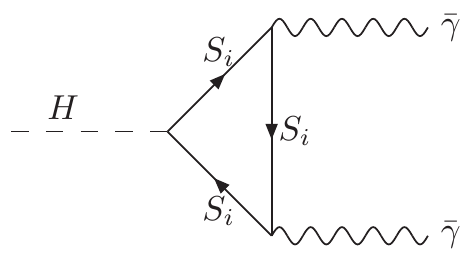}
\vskip 0.5cm
\caption{Higgs decays  $H\to
\gamma\, \bar{\gamma}\, ,\;Z\, \bar{\gamma}\, ,\; \bar\gamma\, \bar{\gamma}\,,$
via mediator loops.}
\label{fig:Hdecays}
\end{center}
\end{figure}

 The Higgs boson can interact with  dark photons  radiatively.
 In the framework proposed in \cite{Gabrielli:2013jka}, this occurs at one loop by the exchange of  scalar messenger fields (Figure~\ref{fig:Hdecays}).

 As a consequence, the Higgs boson can act as a portal toward the dark sector, 
 giving rise to new Higgs-boson decays such as \cite{Gabrielli:2014oya}
\bea
H\to \gamma\, \bar{\gamma}\, ,\;Z\, \bar{\gamma}\, ,\; \bar\gamma\, \bar{\gamma}\,,
\label{dgamma}
\eea
where the symbols $\gamma$ and $\bar{\gamma}$ stand  for 
the usual QED photon and dark photon, respectively, and $Z$ is the neutral vector boson. 
The corresponding decay rates can  in principle be large, even for very heavy messenger fields. 
As in the $H\to \gamma\, \gamma, Z\gamma, gg\;$  decays in the SM, the non-decoupling
Higgs properties guarantee non-vanishing decay widths even in the large-mass
limit for particles exchanged in the loop, provided the virtual (messenger) fields  carry the same $SU(2)_L$ quantum numbers of quarks and leptons.

Being fully decoupled at tree level from the SM sector, a single massless dark photon  will give rise in the Higgs final state to same amount of missing energy and  missing momentum, while the two dark-photon channel will  contribute to the invisible Higgs rate.
Extra contributions to the widths of the SM channels $H\to \gamma\, \gamma, Z\gamma, gg\;$ are also expected in general.

The $H\to \gamma\, \bar{\gamma}\,$ decay  gives rise to a new 
spectacular signature at the LHC in $\gamma + \slash \ET $ final states, with a  photon plus missing transverse energy $\hspace{0.05 cm}\slash \ET$ resonating at the Higgs mass.  
In \cite{Gabrielli:2014oya}, a parton-level study shows that the LHC Run-1  data set 
could be sensitive to BR($H\to\gamma\bar{\gamma}$) values as low as $0.5\%$, while a minimal-model
prediction for BR($H\to\gamma\bar{\gamma}$) can be as large as~$5\%$.

\begin{figure}
\begin{center}
\includegraphics[width=0.35\textwidth]{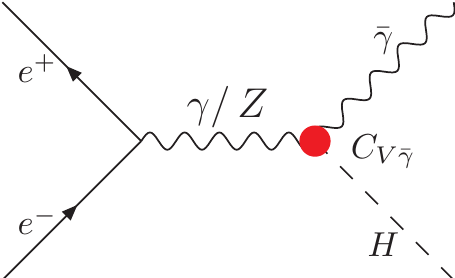}
\caption{Diagrams for  $e^+e^-\to H\bar{\gamma}$.}
\label{fig:prod_diag}
\end{center}
\end{figure}

The aim of the present study is to analyze the phenomenological implications of the Higgs effective couplings to dark photons at future  $e^+e^-$ colliders \cite{Behnke:2013xla, Aicheler:2012bya,FCC}. Apart from the   new signatures corresponding to the Higgs-boson exotic  decays in Eq.~(\ref{dgamma}) (that we do not address here),   effective   $H\bar{\gamma}\gamma$ and $H\bar{\gamma}Z\;$ 
interactions involving dark photons will give rise to  final
states with a Higgs boson and a dark photon, 
\bea
e^+ e^- \to H \; \bar{\gamma} \, ,
\label{eeHdgamma}
\eea
 via $s$-channel exchange of either a photon or a $Z$ vector boson  
 (Figure~\ref{fig:prod_diag}).
This channel, although  kinematically similar to the SM one-loop channel
$e^+e^-\to H {\gamma}$ 
\cite{Barroso:1985et}, 
 gives rise to a completely new signature, since 
the final massless $\bar \gamma$ goes undetected.
 
We will focus on the $b\,\bar b \, \bar{\gamma} $ final state corresponding to the main Higgs decay channel  
$H\to b\bar{b}$,
although even more rare Higgs decays will be of relevance  in the clean
$e^+e^-$ environment  \cite{Baer:2013cma,Asner:2013psa}. 
The $e^+ e^- \to H \bar{\gamma}$ final state will then be characterized
by  an unbalanced $b\,\bar{b}$ system  resonating at the Higgs mass $m_H$, the dark photon $\bar{\gamma}$ giving rise   to
``monochromatic'' missing energy $\slashed{E}$ and momentum $\slashed{p}$
(for fixed initial c.m. collision energy $\sqrt s$). Contrary to what occurs in  the main irreducible SM $b\bar{b}\nu\bar\nu$ background, at parton level 
the invariant mass of the invisible system $M_{miss}=(\slashed{E}^{\,2}-\slashed{p}^{\,2})^{1/2}$  vanishes. This feature will provide a crucial handle for background suppression.

Since the messenger fields are expected to be quite heavy  with respect to the characteristic energy of the   $e^+ e^- \to H \bar{\gamma}$ process, the 
$H\gamma\bar{\gamma}$ and $HZ\bar{\gamma}$  vertices can be considered as effective interactions, and  described by  a model-independent parametrization
\cite{Gabrielli:2014oya}. 
The ratio of the $H\bar{\gamma}Z$
and $H\bar{\gamma}\gamma$ couplings will in general depend on the spin and 
the SM gauge-group representation of the new particles running in the loop.
For simplicity, we will focus here on scenarios where the $H\bar{\gamma}Z$ vertex is induced by scalar messenger fields in the $SU(2)_L\times SU(3)_c$ fundamental representation   \cite{Gabrielli:2013jka}, which gives a definite prediction for the 
$H\bar{\gamma}Z$
and $H\bar{\gamma}\gamma$ coupling ratio.

The paper is organized as follows.  In Section 2, we provide
a model-independent parametrization of the effective couplings
controlling the Higgs exotic decays  
$H\to \gamma\bar{\gamma}, \; Z\bar{\gamma}, \; \bar{\gamma}\bar{\gamma}$, and  
the SM-like decays $H\to \gamma\gamma, \;Z\gamma$, and 
 express the relevant Higgs BR's in terms of the model-independent coefficients.
In Section 3, we study the sensitivity of future $e^+e^-$  colliders to the  
 $e^+e^-\to H\bar\gamma$ associated production by analyzing  the signal  and  corresponding backgrounds. 
In Section 4, we discuss the NP model  in \cite{Gabrielli:2013jka} that aims to solve the Flavor hierarchy  problem. We also present the corresponding predictions for the 
Higgs-dark-photon effective couplings, and   for the 
Higgs branching ratios (BR's) relative to the decays  $H\to \gamma \bar{\gamma}$, and
$H\to Z\bar{\gamma}$. Finally, our conclusions are discussed in Section 5.
In the Appendix, we describe some $U(1)_F$ coupling properties of the model  in \cite{Gabrielli:2013jka},
that are needed to discuss its phenomenological consequences.

\section{Effective dark-photon  couplings to the Higgs boson}
\begin{figure}
\begin{center}
\includegraphics[width=0.45\textwidth]{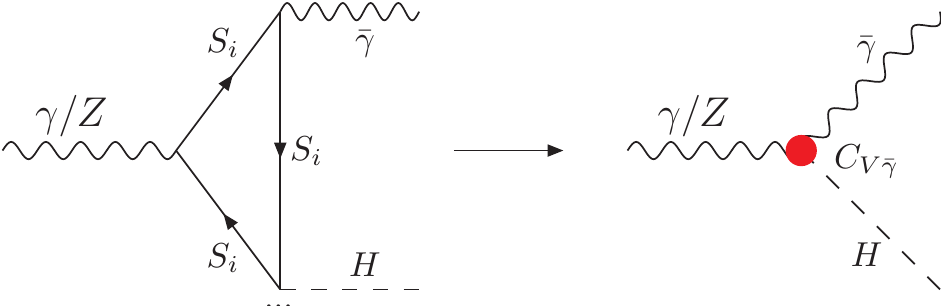}
\caption{Effective coupling approximation for the vertices 
$H
\gamma\, \bar{\gamma}\, ,\;HZ\, \bar{\gamma}$.}
\label{fig:effective}
\end{center}
\end{figure}

We now introduce  the dark-photon effective couplings  to the Higgs boson
that enter the $e^+e^-\to H\bar\gamma$ cross section.
In general, Higgs-dark-photon effective couplings can 
 arise at one loop due to the exchange of  messenger fields that are  charged under both the SM  and the $U(1)_F$ gauge groups (Figure~\ref{fig:effective}). 
In case the messenger masses   are much larger than both 
$m_H$ and  $\sqrt s$,  one can use the effective theory approximation. 
The corresponding effective Lagrangian 
${\cal L}^{\rm Higgs}_{\rm eff}$ can be split as
\bea
{\cal L}^{\rm Higgs}_{\rm eff}&=&{\cal L}_{\rm DP_H}+{\cal L}_{\rm SM_H}\, ,
\label{Ltot}
\eea
where ${\cal L}_{\rm DP_H}$ contains the dark-photon effective interactions  with the Higgs boson, while ${\cal L}_{\rm SM_H}$
presents the {\it extra} (that is messenger-induced) contributions  to the  SM Higgs effective interactions
with two photons, one photon and a $Z$, and two gluons. 

By retaining only the relevant low-energy operators,  ${\cal L}_{\rm DP_H}$  can be expressed in terms of dimensionless (real) coefficients $C_{i\,j}$ (with $i,j =\bar \gamma , {\gamma}, Z, g$) as
\bea
 {\mathcal{L}}_{\rm DP_H} &=& \frac{\alpha}{\pi}\Big(\frac{C_{\gamma \bar{\gamma}}}{v}\gamma^{\mu \nu}\bar{\gamma}_{\mu \nu} H \, +\,  
\frac{C_{Z\bar{\gamma}}}{v} Z^{\mu \nu}\bar{\gamma}_{\mu \nu} H
\, +\, \frac{C_{\bar{\gamma}\bar{\gamma}}}{v} \bar{\gamma}^{\mu \nu}\bar{\gamma}_{\mu \nu} H\Big),
\label{Leff}
\eea
where $\alpha$ is the  SM fine structure constant,
 and $\gamma_{\mu \nu}$, $Z_{\mu \nu}$, $\bar{\gamma}_{\mu \nu}$ are the  field strentghs of photon, $Z$ boson, and dark photon,  respectively ($\gamma_{\mu \nu}\equiv \partial_{\mu} A_{\nu}-\partial_{\nu} A_{\mu}$ for the photon field $A_{\mu}$). Then, ${\cal L}_{\rm SM_H}$ can be written as 
\bea
{\mathcal{L}}_{\rm SM_H}&=& \frac{\alpha}{\pi}\Big(\frac{C_{\gamma \gamma}}{v}\gamma^{\mu \nu}\gamma_{\mu \nu} H \, +\,  
\frac{C_{Z\gamma}}{v} Z^{\mu \nu}\gamma_{\mu \nu} H\Big)
\, +\, \frac{\alpha_S}{\pi}\frac{C_{gg}}{v} G^{a \mu \nu}G^a_{\mu \nu} H,
\label{LeffSM}
\eea
where $\alpha_S$ is the SM strong coupling constant, $G^{a \mu \nu}$ stands for the gluon field strength, and a sum over the color index $a$ is understood. 

As usual, the $C_{i\,j}$ coefficients  in Eqs.(\ref{Leff})-(\ref{LeffSM}) can be  computed 
 in the complete theory by evaluating  
  one-loop amplitudes 
for relevant physical processes, and by matching them with the corresponding results obtained at tree level via the effective Lagrangian in Eq.(\ref{Ltot}). 
In particular, 
in order to express the coefficients $C_{\gamma \bar{\gamma}},\, C_{Z \bar{\gamma}},\, C_{\bar{\gamma}\bar{\gamma}}$ in Eq.(\ref{Leff}) in terms of the fundamental parameters of the model, one can match 
the tree-level widths, based on the  parametrization in  Eq.(\ref{Leff}), for the Higgs decays  $H\to \gamma \bar{\gamma}$, $H\to Z \bar{\gamma}$,  $H\to \bar{\gamma}\bar{\gamma}$, respectively, with 
the corresponding one-loop results 
 computed in the full model (as sketched in Figure~\ref{fig:effective}). 
This will be discussed in Section 4, after introducing a particular NP framework.

On the other hand, one can perform a phenomenological study of the 
$e^+ e^- \to H \bar{\gamma}$ process just on the basis of  the model-independent parametrization in Eq.(\ref{Leff}), which we will do in the next Section.

Before proceeding, we connect the basic $C_{i\,j}$ coefficients in Eq.(\ref{Leff}) to the 
corresponding $H\to i \,j$  decay widths.
The $H\to \gamma \bar{\gamma}$ width has been computed in  \cite{Gabrielli:2014oya},
and, taking into account the parametrization in Eq.(\ref{Leff}), one has
\bea
\Gamma(H\to \gamma \bar{\gamma})&=&\frac{m_H^3 \alpha^2 |C_{\gamma\bar{\gamma}}|^2}{8 \pi^3 v^2}
\, , ~~~~~
\Gamma(H\to gg)\,=\,\frac{m_H^3 \alpha_S^2 |C_{gg}|^2 (N^2_c-1)}{4 \pi^3 v^2}\, ,
\eea
where $N_c=3$ and $\Gamma(H\to gg)$ is understood to be inclusive in gluons final
states.
Analogous results can be obtained for the  $H\to \bar{\gamma}\bar{\gamma}$,
$H\to Z \bar{\gamma}$,   $H\to \gamma \gamma$ widths replacing  
$|C_{\gamma \bar{\gamma}}|^2\;$ by $2|C_{\bar{\gamma} \bar{\gamma}}|^2$,
$|C_{Z \bar{\gamma}}|^2$, $2|C_{\gamma \gamma}|^2$ respectively.

\begin{figure}
\begin{center}
\includegraphics[width=0.6\textwidth]{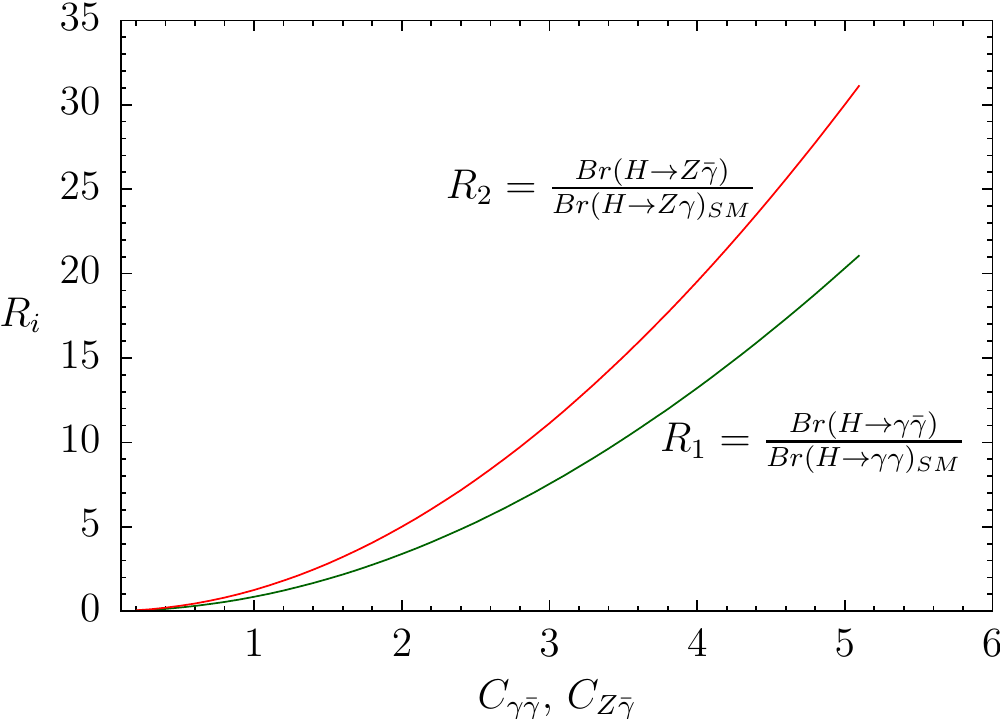}
\caption{BR's for $H\to \gamma \bar{\gamma}$ and 
$H\to Z \bar{\gamma}$, normalized to the SM values for 
BR($H\to \gamma {\gamma}$) and 
BR($H\to Z {\gamma}$), respectively,
  versus $C_{\gamma\bar{\gamma}}$ and $C_{Z\bar{\gamma}}$.}
\label{fig:BRversusC}
\end{center}
\end{figure}

Figure~\ref{fig:BRversusC} shows the  branching ratios for $H\to \gamma \bar{\gamma}$ and 
$H\to Z \bar{\gamma}$, normalized to the SM BR($H\to \gamma {\gamma}$) and 
BR($H\to Z {\gamma}$), respectively,
 versus the corresponding $C_{i\,j}$ coefficients. The $C_{i\,j}$ ranges shown in the plot include values well allowed by the model described in Section 4.  
 One can then  get for the Higgs decays into a dark photon an enhancement factor  
${\cal O}(10)$ with respect  to the SM Higgs decays where the dark  photon is replaced by a photon. This makes the corresponding  phenomenology quite relevant for both LHC and 
future-collider studies.

It is also useful to express  the BR's  for   $H\to \gamma \bar{\gamma},\; \bar \gamma \bar{\gamma}, \;  \gamma {\gamma}  $  as a function of the 
{\it relative} exotic contribution $r_{i\,j}$ to the  $H\to i \,j$ decay width, as 
the ratio   
\bea
r_{ij}&\equiv&\frac{\Gamma^{\rm m}_{i\,j}}{\Gamma^{\rm SM}_{\gamma\gamma}}\; ,
\label{rij}
\eea
with $\Gamma^{\rm m}_{i\ j}$ generically indicating the pure messenger contribution to $H\to  i \,j$, with  $i,j = \gamma , \bar{\gamma}$.

Then, one obtain the following model-independent parametrization
of the  
$H\to {\gamma}\bar{\gamma},\; \bar \gamma \bar{\gamma}, \;  \gamma {\gamma}  $
BR's as functions of $ r_{ij}$ \cite{Gabrielli:2014oya}
\bea
B\!R_{\gamma\bar{\gamma}}&=&B\!R^{\rm SM}_{\gamma\gamma}
\frac{r_{\gamma\bar{\gamma}}}
{1+r_{\bar{\gamma}\bar{\gamma}}B\!R^{\rm SM}_{\gamma\gamma}}\, ,
\nonumber\\
B\!R_{\bar{\gamma}\bar{\gamma}}&=&B\!R^{\rm SM}_{\gamma\gamma}
\frac{r_{\bar{\gamma}\bar{\gamma}}}
{1+r_{\bar{\gamma}\bar{\gamma}}B\!R^{\rm SM}_{\gamma\gamma}}\, ,
\nonumber\\
B\!R_{\gamma\gamma}&=&B\!R^{\rm SM}_{\gamma\gamma}
\frac{\left(1+\chi \sqrt{r_{\gamma\gamma}}\right)^2}
{1+r_{\bar{\gamma}\bar{\gamma}}B\!R^{\rm SM}_{\gamma\gamma}}\, ,
\label{BRS}
\eea
where  $\chi=\pm 1$ parametrizes the relative sign of the SM and exotic amplitudes,
and ${\rm BR}_{i j} $ stands for  ${\rm BR}(H\to i\, j) $.

Analogously,  the {\it relative} deviation for the $H\to gg$ decay width will be defined as
\bea
r_{gg}&\equiv&\frac{\Gamma^{\rm m}_{gg}}{\Gamma^{\rm SM}_{gg}}\, .
\label{rgg}
\eea

\section{Sensitivity study for  $e^+e^-\to H \bar{\gamma}$}

\begin{figure}
\begin{center}
\includegraphics[width=0.55\textwidth]{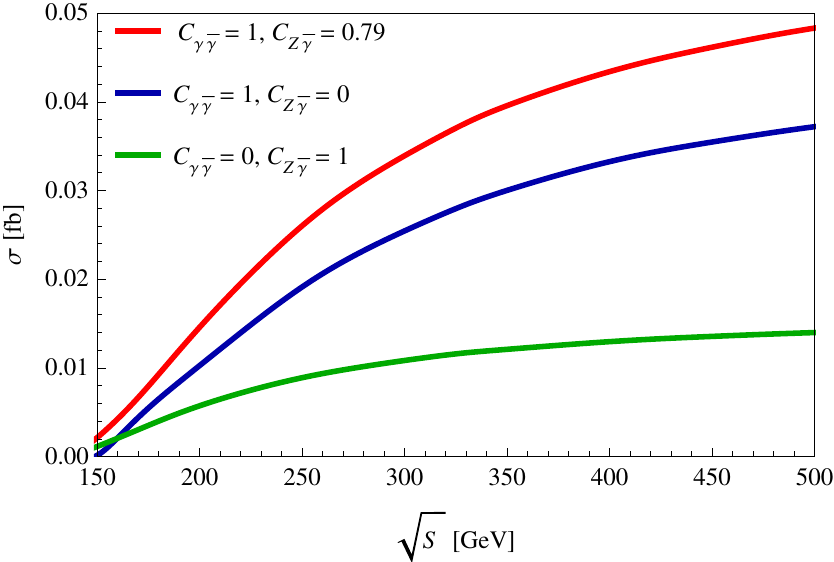}
\caption{Total $e^+e^-\to H\bar{\gamma}\,$ cross section  as a function of the c.m. collision energy, for different sets of effective couplings.}
\label{fig:xsection-energy}
\end{center}
\end{figure}

We focus
 now on the $\bar{\gamma}$ production in association with a Higgs boson in $e^+e^-$
 collisions. The \mbox{$e^+e^-\to H \bar{\gamma}$} total cross section  versus $\sqrt s$ is shown in figure \ref{fig:xsection-energy} 
 for three different coupling assumptions: $C_{\gamma\bar{\gamma}}=1,C_{Z\bar{\gamma}}=0$ (blue line);
 $C_{\gamma\bar{\gamma}}=0,C_{Z\bar{\gamma}}=1$ (green line);
 $C_{\gamma\bar{\gamma}}=1, C_{Z\bar{\gamma}}=0.79 \;C_{\gamma\bar{\gamma}}$ (red line).  
 The coupling ratio $C_{Z\bar{\gamma}}/C_{\gamma\bar{\gamma}}\simeq 0.79 $ is typical for scenarios where 
the $H\bar{\gamma}Z$ vertex is induced by scalar messenger fields in the $SU(2)_L\times SU(3)_c$ fundamental representation (see Section 4).
The corresponding cross sections at $\sqrt s\simeq1$ TeV (relevant for linear colliders at larger collision energy) are  43 ab, 15 ab, 55 ab,
 respectively.
 Cross sections can be  easily extrapolated to coupling set-up obtained just by globally rescaling these set of couplings.
 
 The  \mbox{$e^+e^-\to H \bar{\gamma}$} cross sections  grow with  c.m. energy thanks to the nature of the  
 dimension-five operators in the effective Lagrangian in Eq.~(\ref{Leff}).
 Hence, at constant integrated luminosity, higher-energy colliders will have a higher potential, since the dominant background is expected to scale down with energy
 as $1/s$. On the other hand, lower $\sqrt s$ may allow larger integrated luminosity,
 as is the case of the $e^+e^-$ Future Circular Collider (FCC-ee) (also called TLEP) \cite{FCC},
 where an integrated luminosity of 10 ab$^{-1}$ is expected at $\sqrt s=240$~GeV.
 At linear colliders, either ILC \cite{Behnke:2013xla} or CLIC \cite{Aicheler:2012bya}, one typically foresees integrated luminosities of a few hundreds
fb$^{-1}$  in the initial energy of
$\sqrt{s} \sim$ 250 GeV or 350 GeV, and a few ab$^{-1}$ at the larger$-\sqrt{s}$ 
stages \cite{Dawson:2013bba}.
 Here, we assume the minimal energy setup of $\sqrt s=240$~GeV that is relevant for Higgs-boson studies, and  study the sensitivity to $e^+e^-\to H \bar{\gamma}$ production 
 versus integrated luminosities foreseen at different machines.

Using the effective Lagrangian Eq.~(\ref{Leff}) implemented by FeynRules 
\cite{Alloul:2013bka}, we have generated \mbox{$e^+e^-\to H \bar{\gamma}\to b\bar{b}\bar{\gamma}$} events with MadGraph5$\_$aMC@NLO  \cite{Alwall:2014hca}, and passed these events to PYTHIA to account for parton showering, and hadronization.
We checked that the inclusion of effects from initial state radiation, that tends to degrade the c.m. energy in a circular $e^+e^-$ colliders, would moderately affect  the results of the present analysis. We neither  
include beamstrahlung  effects that can be of some relevance at linear colliders.
 We account for finite detector resolution by applying the 
jet-energy smearing  $\sigma(E)/E = 30\%/\sqrt{E}$, which is typical for ILC-kind of detectors  \cite{Behnke:2013lya} .

The dark photon escapes the experimental apparatus undetected, and  the final signal consists of two $b$ quarks and large missing energy $\slashed{E}$ and momentum
$\slashed{p}$. In our simulation we reconstruct the missing momentum from the vector sum of all visible final-state particle momenta, after applying PYTHIA. In a lepton collider a  $H\to b\,\bar{b}$ final state is not swamped by  large QCD backgrounds, as occurs in hadronic collisions. Therefore,  $b\,\bar{b}$ final states are the best channel to search for $H\bar{\gamma}$ production, thanks to the  $H\to b\,\bar{b}$ large rate. 
After showering and hadronization, we reconstruct jets (and $b$-jets) according to the basic PYTHIA jet-cone algorithm, assuming a quite large cone aperture  $R_{j}=1.5$, which optimizes mass reconstruction \cite{Seidel:2013sqa}. The basic  event selection is given by
\begin{eqnarray}
\begin{tabular}{l}
$p_{T}^{b}>20$~GeV\;,\;\;\;
$|\eta_{b}|<2.5$\;,\;\;\;
$\Delta R(bb) > 0.4$\;,\;\;\;
$\slashed{E}>40$~GeV,
\end{tabular}
\label{eq:eventselection}
\end{eqnarray}
where $\Delta R(bb)=\sqrt{\Delta \eta^2+\Delta \phi^2}$ is the angular distance between two $b$-tagged jets.
We assume a $b$-tagging efficiency of 80\%, and a corresponding fake $b$-jet rejection factor of 100 for light jets.

The main SM background for the $b\,\bar{b}+\slashed{E}$ final state is given by the $\nu\bar{\nu}b\bar{b}$ production. This includes the on-shell processes $ZZ \to \nu\bar{\nu}b\bar{b}$, $ZH \to \nu\bar{\nu}b\bar{b}$, which give an almost monochromatic $b\bar{b}$-pair 
 system (similarly to the signal), and the vector boson
fusion channel $H\nu\bar{\nu}$. A subdominant contribution comes from $\nu\bar{\nu}q\bar{q}$ (mostly from on-shell $Z$ pairs), where both light jets are mis-tagged as $b$ jets.

\begin{figure}
\begin{center}
\includegraphics[width=0.6\textwidth]{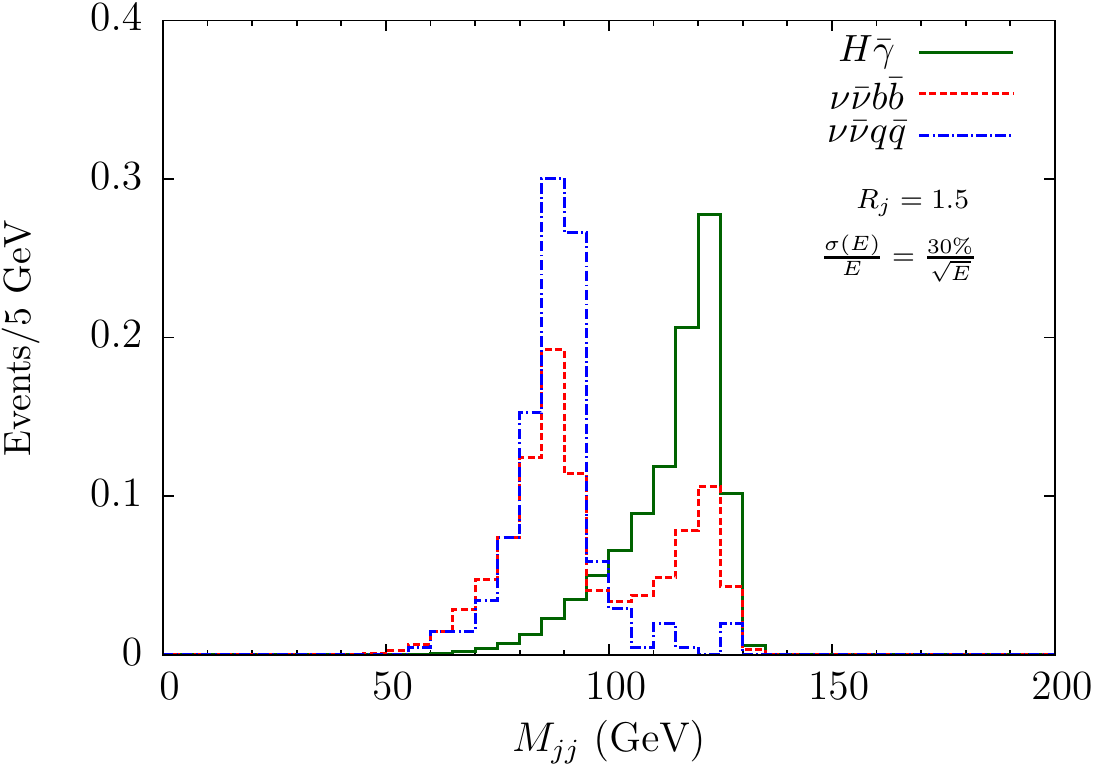}
\caption{Invariant-mass $M_{jj}$ distributions for the two  jets with largest $p_T$ for the
the signal (solid line) and the two backgrounds $\nu\bar{\nu}b\bar{b}$ (dashed line) and 
$\nu\bar{\nu}q\bar{q}$ (dot-dashed line) after PYTHIA showering, hadronization, and 
jet-energy resolution effect. All distributions are normalized to 1.}
\label{fig:Mjj}
\end{center}
\end{figure}

There are two kinematical variables that turn out to be particularly efficient  in  separating the signal from the background. First, 
we introduce  the variable $M_{jj}$ as the  invariant mass of  the two  jets with largest $p_T$. This is directly connected to 
the $b$-pair invariant mass, and   can be used to pinpoint events with  $b$-quarks coming from Higgs decays, out of  the smaller-$M_{jj}$  events arising from $Z\to b\bar{b}$. There is anyway  part of the $\nu\bar{\nu}b\bar{b}$   background 
that goes through the $ZH$ production resonating at $M_{jj}\sim m_H$, just as in the signal case. This is well illustrated by Figure~\ref{fig:Mjj}, where  the normalized invariant-mass distributions of the $b\bar{b}$ system  are compared for  signal and backgrounds. 
Second, we introduce the  {\it missing-mass} variable $M_{\rm miss}$,  defined as
\be
M_{\rm miss} = \sqrt{{\rm \slashed{E}^2}-{\rm \slashed{\bf p}}^2},
\ee
where ${\rm \slashed{E}}=\sqrt s-\sum {\rm E}_{visible}$ and ${\rm \slashed{\bf p}}=-\sum {\mathbf p}_{visible}$ are the final-state missing energy and missing three-momentum vector, respectively (the sum over visible objects here includes both jets and lower-energy  particles escaping  jet reconstruction).
The $M_{\rm miss}$ variable is expected to approximately vanish in the partonic description of $e^+e^-\to H \bar{\gamma}$,
corresponding to the massless 
 {\it invisible} dark photon. 
 A cut on $M_{\rm miss}$ then proves to be remarkably efficient in further separating the signal from the main 
 background, where $M_{\rm miss}$ mostly matches  an invisible $Z$-boson decaying into neutrinos. 
 
\begin{figure}
\begin{center}
\includegraphics[width=0.48\textwidth]{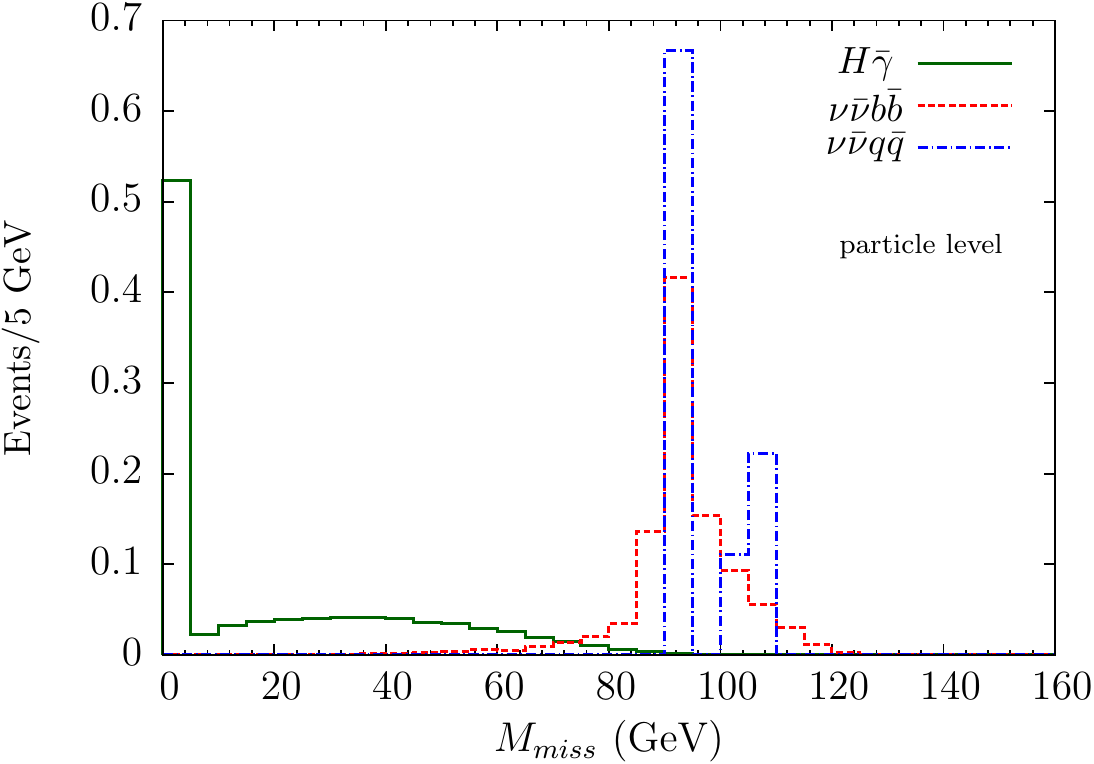}
\includegraphics[width=0.48\textwidth]{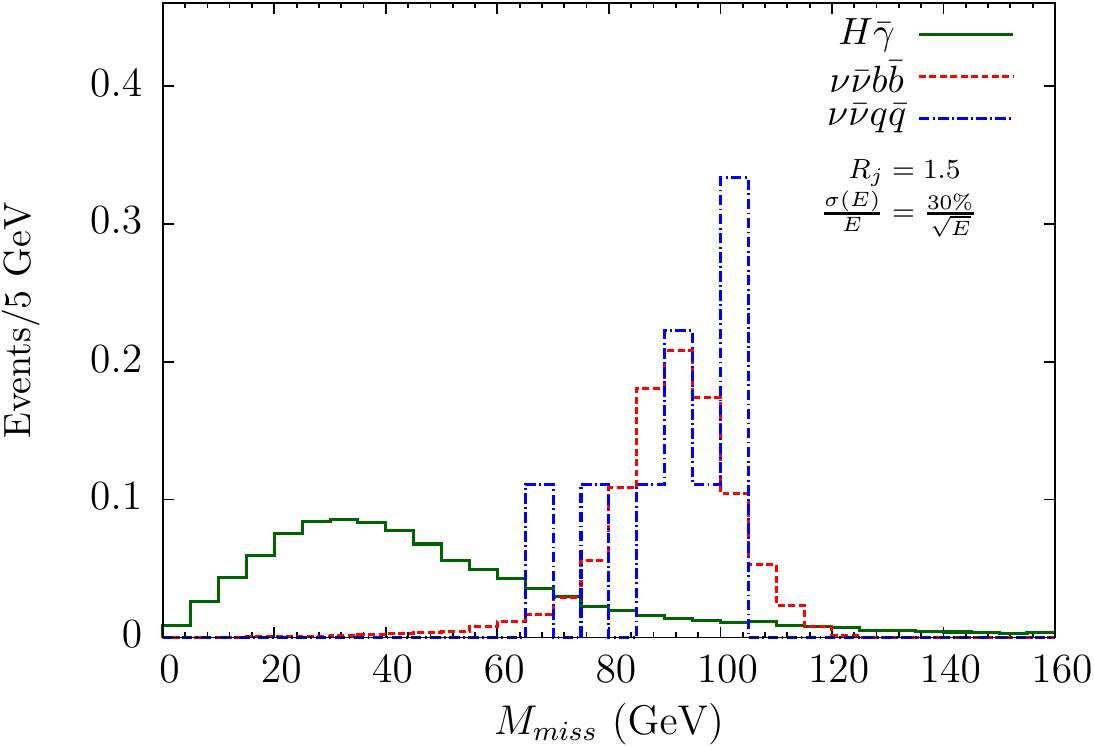}
\caption{$M_{\rm miss}$ distributions for the
the signal (solid line) and the backgrounds $\nu\bar{\nu}b\bar{b}$ (dashed line) and 
$\nu\bar{\nu}q\bar{q}$ (dot-dashed line) after parton level simulation (left) and after PYTHIA showering, hadronization and 
jet energy resolution effect (right). All distributions are normalized to~1.}
\label{fig:missingmass}
\end{center}
\end{figure}

 The $M_{\rm miss}$ spectrum of the signal and background processes are compared in Figure~\ref{fig:missingmass}, after  applying PYTHIA showering, jet reconstruction and 
 jet-energy resolution effects on top of parton-level simulation (right panel). 
 The parton-level spectrum, shown in the left panel of the same figure, shows a distinct peak at $M_{\rm miss}\simeq0$ for the signal, and at $M_{\rm miss}\sim M_Z$ for the background processes. No energy-resolution effect has been applied in the latter case, and the smearing of the peaks is just due to the presence of  neutrinos from $b$ decays, and to  the possible  off-shellness of the $\nu\bar\nu$ system in the  background.  Applying the parton showering, jet reconstruction and energy-resolution effects (as in the right panel of the figure) degrades the $M_{\rm miss}$ spectrum of the signal quite a lot, shifting the peak away from zero and smearing it. Hence, an optimal detector resolution would be  particularly crucial in this analysis.

\begin{figure}
\begin{center}
\includegraphics[width=0.6\textwidth]{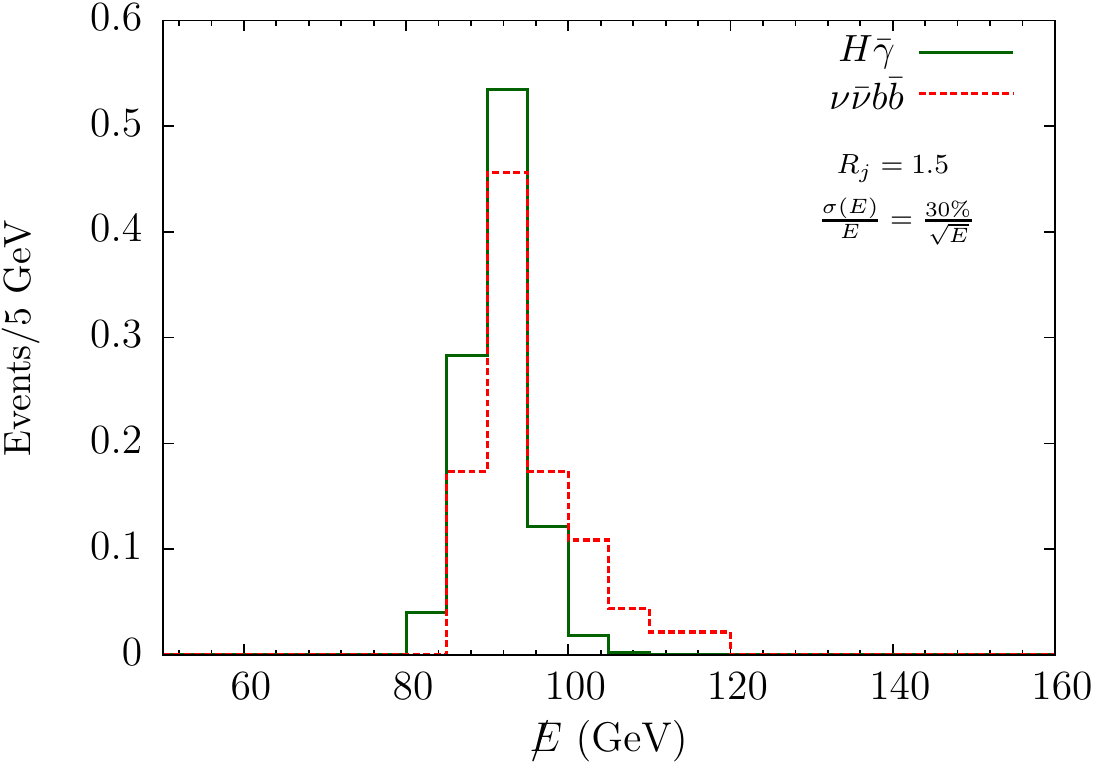}
\caption{$\slashed{E}$ distributions for the
the signal (solid line) and the background $\nu\bar{\nu}b\bar{b}$ (dashed line) after PYTHIA showering, hadronization and jet-energy resolution effect, and after applying the 
$M_{jj}$ and $M_{\rm miss}$ cuts described in the text. All distributions are normalized to 1.}
\label{fig:ME}
\end{center}
\end{figure}

On the basis of the $M_{jj}$ and $M_{miss}$ distributions  in Figures \ref{fig:Mjj} and \ref{fig:missingmass}, we set a suitable event selection. We require the invariant mass $M_{jj}$ to be within 10\% of the $M_{jj}$ peak value  of the simulated signal events, and then impose the missing mass to be below 40 GeV. The latter cuts make the $\nu\bar{\nu}q\bar{q}$ background negligible. The $\nu\bar{\nu}b\bar{b}$ background can still be slightly reduced after these cuts by making a further cut on the missing energy $\slashed{E}$. The $\slashed{E}$ spectrum is shown in Figure \ref{fig:ME} for the signal and background events satisfying the previous $M_{jj}$ and $M_{\rm miss}$ cuts. Both the signal and background distributions peak at around the same value, with the background  moderately shifted to larger $\slashed{E}$ values. Thus we require the missing energy to be below 100 GeV. Including the initial event selection criteria, we altogether
impose  that the missing energy satisfies the condition \mbox{$40\ {\rm GeV} < \slashed{E} < 100\ {\rm GeV}$.}

Table \ref{table:acceptancies} shows the cross sections and the acceptances for the signal and the $\nu\bar{\nu}b\bar{b}$ background after applying  the cut-flow just described, 
for \mbox{$\sqrt s=$240 GeV.} The signal acceptance is practically insensitive to
to a change in the relative contribution of the $C_{\gamma \bar{\gamma}}$ and 
$C_{Z \bar{\gamma}}$ couplings.
The corresponding acceptance for the $\nu\bar{\nu}q\bar{q}$ background is  negligible.

\begin{table}
\begin{center}
\begin{tabular}{||l||c||c||}
\hline 
\hline
Process & Cross section (fb) & Acceptance after cuts (\%) \\
\hline 
$ H \bar{\gamma}$\;\; ($C_{Z \bar{\gamma}}= 0$)   &  $10.1\times 10^{-3} \;C^2_{\gamma \bar{\gamma}}$   & 17.3  \\
\hline
$ H \bar{\gamma}$\;\; ($C_{\gamma \bar{\gamma}}= 0$)     &  $4.8\times 10^{-3}\;C^2_{Z \bar{\gamma}}$   & 17.3  \\
\hline
$ H \bar{\gamma}$\;\; ($C_{Z \bar{\gamma}}=0.79 \;C_{\gamma \bar{\gamma}}$)   &  $13.8\times 10^{-3} \;C^2_{\gamma \bar{\gamma}}$   & 17.3 \\
\hline
SM $\;\nu\bar{\nu}b\bar{b}$   &  115.   &  0.08  \\
\hline
\hline
\end{tabular}
\caption{ Cross sections (in fb) and corresponding acceptances after kinematical  cuts on signal and SM background at  \mbox{$\sqrt s=$240 GeV.}  Applied cuts include the initial event selection in Eq.~(\ref{eq:eventselection}), 
$M_{jj}$ to be within 10\% of the $M_{jj}$ peak value  of  signal events, \mbox{$M_{\rm miss}<40$ GeV}, and \mbox{$\slashed E < 100$ GeV. Cross sections include BR$(H\to
b\bar{b})\simeq0.58$.}  } 
\label{table:acceptancies}
\end{center}
\end{table}      

\begin{figure}
\begin{center}
\includegraphics[width=0.7\textwidth]{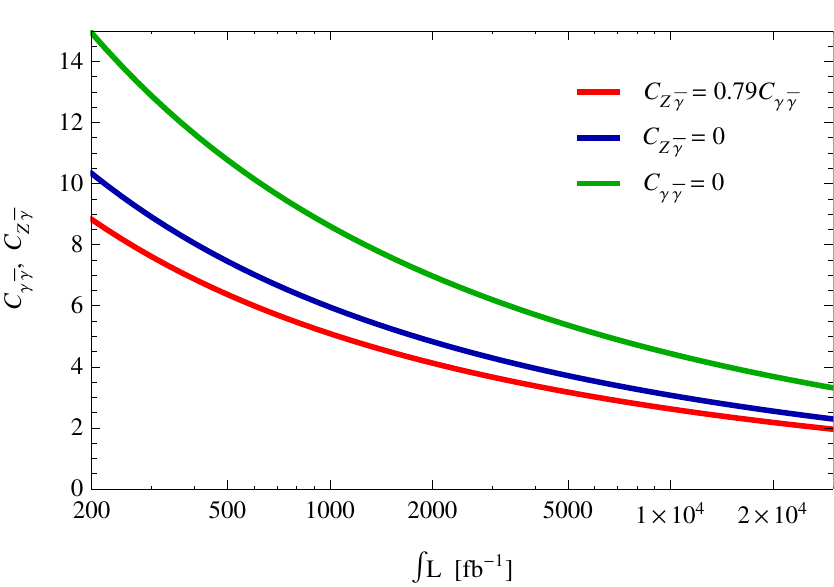}
\caption{ The $5\,\sigma$-sensitivity bounds for $e^+ e^- \to H \bar{\gamma}$  as a function of the integrated luminosity
at $\sqrt s = 240$ GeV, when $C_{\gamma\bar{\gamma}}=0$ (green), $C_{Z\bar{\gamma}}=0$ (blue), and $C_{Z\bar{\gamma}}=0.79\,C_{\gamma\bar{\gamma}}$ [with  $C_{\gamma\bar{\gamma}}$ shown on the vertical axis]  (red).}
\label{fig:reach}
\end{center}
\end{figure}

On the basis of  the Table \ref{table:acceptancies} acceptances, we can work out the expected sensitivity to the signal for given values of the  $C_{\gamma\bar{\gamma}},C_{Z\bar{\gamma}}$ couplings.
As usual, we define the signal significance as $S/\sqrt{S+B}$, being $S$ and $B$ the
event numbers for signal and background, respectively.  Figure~\ref{fig:reach} shows the integrated luminosity
needed to make a $5\, \sigma$ observation of the $ H \bar{\gamma}$ production in $e^+e^-$
collisions at  \mbox{$\sqrt s=$240 GeV}, for any given value of the 
$C_{\gamma\bar{\gamma}},C_{Z\bar{\gamma}}$ couplings (shown on the $y$-axis) when
$C_{\gamma\bar{\gamma}}=0$ (green line), $C_{Z\bar{\gamma}}=0$ (blue line) and $C_{Z\bar{\gamma}}=0.79\,C_{\gamma\bar{\gamma}}$ (red line).

\begin{figure}
\begin{center}
\includegraphics[width=0.7\textwidth]{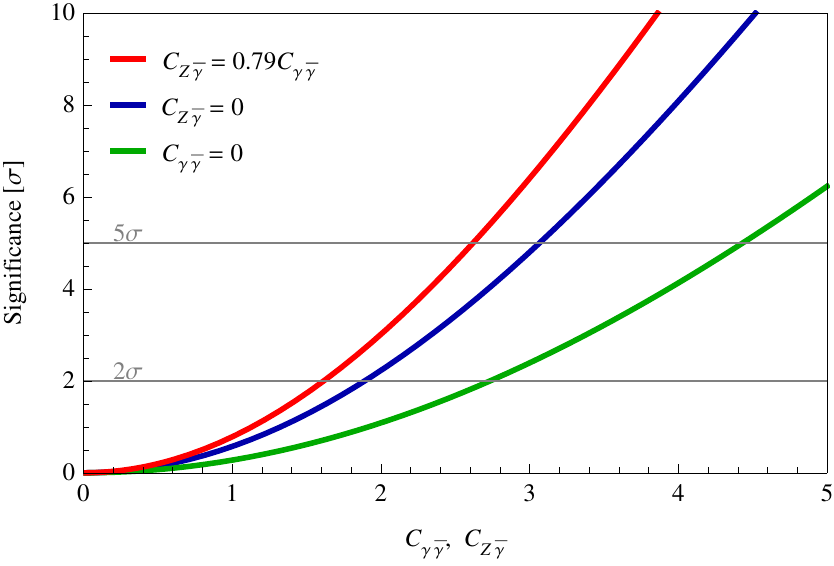}
\caption{Signal significance $S/\sqrt{S+B}\;$ for $e^+ e^- \to H \bar{\gamma}\,$ as a function of the couplings $C_{\gamma\bar{\gamma}},C_{Z\bar{\gamma}}$, for $C_{\gamma\bar{\gamma}}=0$ (green), $C_{Z\bar{\gamma}}=0$ (blue), and $C_{Z\bar{\gamma}}=0.79\;C_{\gamma\bar{\gamma}}\;$ [with  $C_{\gamma\bar{\gamma}}$ shown on the horizontal axis]  (red), for an integrated luminosity of \mbox{$10\ {\rm ab}^{-1}$}, at  \mbox{$\sqrt s=$240 GeV.} The horizontal gray lines show the 
$5\,\sigma$-discovery bound, and  the $2\,\sigma$  ($\simeq 95\%$ C.L. exclusion) level.}
\label{fig:significance}
\end{center}
\end{figure}

For an integrated luminosity of \mbox{$10\ {\rm ab}^{-1}$} at $\sqrt s = 240$ GeV (a typical value for FCC-ee), Figure \ref{fig:significance} shows the signal 
 significance  as a function of the couplings, with the same color convention as in Figure \ref{fig:reach}. The horizontal gray lines show the 
$5\,\sigma$-discovery  bound on couplings, and  the $2\,\sigma$ level  approximating the 95\% confidence-level exclusion.

Then, at 95\% C.L., 
one can exclude  the ranges $C_{\gamma\bar{\gamma}}>1.9$
(for $C_{Z\bar{\gamma}}=0$), $C_{Z\bar{\gamma}}>2.7 $ (for $C_{\gamma\bar{\gamma}}=0$), and $C_{\gamma\bar{\gamma}}>1.6$ (for
$C_{Z\bar{\gamma}}=0.79\;C_{\gamma\bar{\gamma}})$.
The interval $C_{\gamma\bar{\gamma}}>1.9$ corresponds to a Higgs BR
into $\gamma\bar\gamma$ that is more than 3 times the SM BR($H\to \gamma {\gamma}$), while $C_{Z\bar{\gamma}}>2.7 $ 
corresponds to a Higgs BR
into $Z\bar\gamma$ that is more than 9 times the SM BR($H\to Z {\gamma}$).

The corresponding sensitivities on the $C_{\gamma\bar{\gamma}}$ and $C_{Z\bar{\gamma}}$ couplings
at the ILC (foreseeing an initial {$\sqrt s=$250 GeV phase, with a typical integrated luminosity of  {$250\ {\rm fb}^{-1}$}) can be estimated from Figure~\ref{fig:reach},  and are about a factor 3 lower than the FCC-ee ones. The latter match   
sensitivities on the corresponding BR($H\to \gamma {\bar\gamma}$) and 
BR($H\to Z {\bar\gamma}$) that are smaller than the FCC-ee ones by 
an order of magnitude.


\section{A model of Flavor with Dark Photons}
In this section, we review the main  aspects of the model proposed
in \cite{Gabrielli:2013jka}, that provides a theoretical framework for the effective description given by the Lagrangian in 
Eq.~(\ref{Ltot}).
Correspondingly, we will obtain  predictions for the Higgs BR's for $H\to \gamma \bar{\gamma}$ and $H\to Z \bar{\gamma}$ (and relevant effective couplings) in terms of the fundamental parameters of the model.

\subsection{The Lagragian}
\label{model}
The basic assumptions of the model introduced  
in \cite{Gabrielli:2013jka} are the following.
 For each SM fermion there exists a dark fermion with same flavor in a hidden sector, that is an heavier fermion replica which is a singlet under the SM gauge group. The chiral- and flavor-symmetry spontaneus breaking is realized in the dark sector as described in the following,
 and communicated to the SM fermions via  renormalizable and flavor universal interactions mediated by  messenger fields.
The Yukawa couplings are radiatively generated, preserving approximately the same flavor hierarchy structure of the dark-fermion sector. 
A similar approach has been proposed in \cite{Ma:2014rua,Fraser:2014ija}, although
in this case the dynamics responsible of the dark-fermion spectrum has not been discussed.

A simple choice for the messenger sector consists of a set of 
{\it scalar} messenger fields that, due to gauge invariance, have to be charged under the SM gauge group with the same quantum numbers as quarks and leptons.   The relevant Lagrangian is \cite{Gabrielli:2013jka} 
\bea
{\cal L}&= {\cal L}^{ Y=0}_{SM} + {\cal L}_{mes} + {\cal L}_{DS},
\label{LagTOT}
\eea
where  ${\cal L}^{Y=0}_{SM}$ stands for the SM Lagrangian without 
tree-level Higgs Yukawa couplings, ${\cal L}_{mes}$ is the Lagrangian 
containing the messenger sector with its couplings to the SM and dark fields,
and ${\cal L}_{DS}$ is the dark-sector Lagrangian including  the 
 dynamics responsible of the flavor hierarchy. 

Dark fermions acquire an exponentially-spread mass spectrum 
 from  non-perturbative dynamical effects. In fact, 
the Lagrangian ${\cal L}_{DS}$, containing  dark fermions and a dark photon,  is given by 
\cite{Gabrielli:2013jka}, 
\bea
{\cal L}_{DS}&=& 
i\sum_i \left( 
\bar{Q}^{U_i}{\cal D}_{\mu}\gamma^{\mu} Q^{U_i}+\bar{Q}^{D_i}{\cal D}_{\mu}\gamma^{\mu} Q^{D_i}\right)
\nonumber \\
&-&
\frac{1}{4} F_{\mu\nu} F^{\mu\nu} + \frac{1}{2\Lambda^2}  \partial^{\mu} F_{\mu\alpha} \partial_{\nu} F^{\nu\alpha},
\label{LagDS}
\eea
where $Q^{U_i}$, $Q^{D_i}$ are the dark fermion fields, partners in the hidden sector of the up ($U_i$) and down ($D_i$) quarks, ${\cal D}_{\mu}=\partial_{\mu}+i g \hat{Q} \bar{A}_{\mu}$ is the covariant derivative related to
the $U(1)_F$ gauge field $\bar{A}_{\mu}$ which is associated to the dark-photon, with $F_{\mu\alpha}$ the corresponding field strength, and $\hat{Q}$  the charge operator acting on the dark fermion fields.  ${\cal L}_{DS}$ can be 
extended to include also the SM leptonic sector in a straightforward way.

The last term in ${\cal L}_{DS}$, involving only the $U(1)_F$ gauge sector,  corresponds to the so-called Lee-Wick term.
It is a higher-derivative term, and the $\Lambda$
scale can be  interpreted as the mass of the associated massive ghost\footnote{
According to the Lee-Wick argument \cite{Lee:1971ix}-\cite{Cutkosky:1969fq}, the presence of a massive ghost field in the spectrum does not spoil unitarity, provided the massive ghost has a finite decay width, which is  
 automatically satisfied in the present scenario.}.
This term  can trigger ChSB, and generate a mass spectrum  non-perturbatively \cite{Gabrielli:2007cp}, by means of the Nambu-Jona-Lasinio mechanism \cite{Nambu:1961tp}. The following Dirac-fermion mass spectrum can be induced on the true vacuum \cite{Gabrielli:2007cp}
\bea
M_{Q_i}=\Lambda \exp\left\{-\frac{2\pi}{3\,\bar{\alpha}(\Lambda) q_i^2} +\frac{1}{4}\right\}\, ,
\label{mgap2}
\eea
where $q_i$ is the $U(1)_F$  charge of the Dirac fermion, and $\bar{\alpha}(\Lambda)$ the corresponding fine structure constant  at the  $\Lambda$ energy scale. This solution is manifestely non-perturbative as can be seen from the 
$\bar{\alpha}$ dependence in $M_{Q_i}$.

The  Lagrangian ${\cal L}_{mes}$  in Eq.~(\ref{LagTOT}) contains the messenger scalar  fields,
\bea
{\cal L}_{mes}&=&{\cal L}^{\rm 0}_{mes}+{\cal L}^{\rm I}_{mes}\,,
\label{MSMS}
\eea
where  ${\cal L}^{\rm 0}_{mes}$ is the kinetic Lagrangian for the 
messenger fields interacting with the SM gauge bosons, while ${\cal L}^{\rm I}_{mes}$ contains the messenger interactions with the dark fermions and the Higgs boson, which  give rise  to the effective Yukawa couplings.

The SM quark quantum numbers set the minimal matter in the 
  messenger sector, which is given by
\begin{itemize}
\item $2N_f$ complex scalar $SU(2)_L$ doublets: $\hat{S}_L^{\U_i}$ and $\hat{S}_L^{\D_i}$,
\item $2N_f$ complex scalar  $SU(2)_L$ singlets: $S_R^{\U_i}$ and $S_R^{\D_i}$,
\item one real $SU(2)_L\times U(1)_Y$ singlet scalar: $S_0$,
\end{itemize} 
where
$\hat{S}_L^{\U_i(\D_i)}=\left(\begin{array}{c}S^{\U_i(\D_i)}_{L_1}\\S^{\U_i(\D_i)}_{L_2}
\end{array}\right)$, 
$N_f=3$, and $i \;(=1,2,3)$ stands for the flavor index for three fermion generations.
The $\hat{S}_{L}^{\U_i,\D_i}$ and  $S_{R}^{\U_i,\D_i}$ fields 
carry the SM quark quantum numbers, and the labels $L,R$ corresponds to the   chirality of the associated SM fermions. 
They couple  to  the EW gauge bosons and to the gluons, as do squarks in the minimal supersymmetric extensions of the SM.  Note that a minimal flavor violation  would require this Lagrangian to be invariant under $SU(N_F)$, where $N_F=2 N_f$  is  the number of
flavors.  

 The messenger mass structure can be described by  four free 
universal mass terms in both the $\hat{S}_{L,R}^{\U_i}$ and $\hat{S}_{L,R}^{\D_i}$ sectors. Note that an even more minimal hypothesis  of  a  common scalar mass for the $L$ and $R$ scalar sectors is also phenomenologically acceptable.

The Lagrangian   ${\cal L}^I_{mes}$ for the 
 messenger interactions with  quarks and  SM Higgs boson  is
\bea
{\cal L}^I_{mes} &=&
g_L\left( \sum_{i=1}^{N_f}\left[\bar{q}^i_L Q_R^{\U_i}\right] \hat{S}^{\U_i}_{L} +
\sum_{i=1}^{N_f}\left[\bar{q}^i_L Q_R^{\D_i}\right] \hat{S}^{D_i}_{L}\right)+
\nonumber\\
&+&
g_R\left(\sum_{i=1}^{N_f}\left[\bar{\scriptstyle U}^i_R Q_L^{\U_i}\right] S^{\U_i}_{R} +
\sum_{i=1}^{N_f}\left[\bar{\scriptstyle D}^i_R Q_L^{\D_i}\right] S^{\D_i}_{R}\right) +
\nonumber\\
&+&
\lambda_S S_0 \left(\tilde{H}^{\dag} S^{\U_i}_L S^{\U_i\dag}_R+ H^{\dag} S^{\D_i}_L S^{\D_i\dag}_R\right)
+ h.c.  + V(S_0),
\label{LagMS}
\eea 
where   
$S_0$ is a real singlet scalar,  $V(S_0)$ its potential, 
$q^i_L$,  ${\scriptstyle U}^i_R$, ${\scriptstyle D}^i_R$, 
stand for the SM fermions, and  $H$ is the SM Higgs  doublet, with
$\tilde{H}=i\sigma_2 H^{\star}$. Contractions on color indices are understood. The two constants $g_L$ and $g_R\,$ are  flavor-universal free parameters that are allowed to  be in the perturbative region $g_{L,R}< 1$.

The  Lagrangian for  the interaction of messenger scalars 
with the SM gauge bosons follows from the universal properties of gauge interactions. We  stress  that the messenger fields carry the same $U(1)_F$ charges as the corresponding dark fermions.

The messenger and dark-fermion quantum numbers are shown in Table~\ref{tab1}
(detailed in \cite{Gabrielli:2013jka}).
\begin{table} \begin{center}    
\begin{tabular}{|c||c|c|c|c|c|}
\hline 
Fields 
& Spin
& $SU(2)_L$ 
& $U(1)_Y$
& $SU(3)_c$
& $U(1)_F$
\\ \hline 
$\hat{S}_L^{\D_i}$
& 0
& 1/2
& 1/3
& 3
& -$\qdi$
\\ \hline
$\hat{S}_L^{\U_i}$
& 0
& 1/2
& 1/3
& 3
& -$\qui$
\\ \hline
$S_R^{\D_i}$
& 0
& 0
& -2/3
& 3
& -$\qdi$
\\ \hline
$S_R^{\U_i}$
& 0
& 0
& 4/3
& 3
& -$\qui$
\\ \hline
$Q^{\D_i}$
& 1/2
& 0
& 0
& 0
& $\qdi$
\\ \hline
$Q^{\U_i}$
& 1/2
& 0
& 0
& 0
& $\qui$
\\ \hline
$S_0$
& 0
& 0
& 0
& 0
& 0
\\ \hline \end{tabular} 
\caption[]{
Spin and gauge quantum numbers for the strongly-interacting messenger fields and 
for dark fermions.   
$U(1)_F$ is the gauge symmetry  in the dark sector.
}
\label{tab1}
\end{center} \end{table}
Finally, after the spontaneous symmetry breaking of the discrete $H\to -H$ parity
symmetry that prevents tree-level Yukawa couplings, all Yukawa couplings are generated at  one loop, and are  finite at any order in perturbation theory \cite{Gabrielli:2013jka,Ma:2014rua,Fraser:2014ija}).
 Assuming almost degenerate  diagonal  messenger masses in the L and R sectors, one finds, from the one-loop vertex computation, that the  effective  Yukawa coupling associated to the quark $i$ is \cite{Gabrielli:2013jka}
\bea
Y^i&=&Y_0(x_i) \exp{\left(-\frac{2\pi}{3\bar{\alpha}(\Lambda) q_i^2}\right)}\, ,
\label{Yeff}
\eea
where  the dark-fermion masses $M_{Q_i}$ have been replaced by Eq.(\ref{mgap2}),
and 
 the one-loop function $Y_0(x_i)$ is given by
\bea
Y_0(x_i)&=&\left(\frac{g_L g_R }{16 \pi^2 }\right)
\left(\frac{\mu_S\Lambda}{\bar{m}^2}\right) C_0(x_i),
\label{Yuk}
\eea
with $\mu_S\equiv \lambda_S \langle S \rangle $, and 
$x_i=M_{Q_i}^2/\bar{m}^2$. Also, 
$\bar{m}$ is the average mass of the messenger fields running in the loop, and 
$C_0(x)=(1-x\left(1-\log{x}\right))/(1-x)^2$.\footnote{
The above results  hold for diagonal Yukawa
couplings. They  
can be easily generalized to include the Cabibbo-Kobayashi-Maskawa (CKM) matrix entering  the charged weak interactions as explained in \cite{Ma:2014rua}.}

\subsection{BR($H\to \gamma\bar{\gamma}$) predictions}
The $C_{\gamma \gamma}$,  $C_{\gamma \bar{\gamma}}$, and $ C_{\bar{\gamma}\bar{\gamma}}$ coefficients  entering the effective Lagrangian in Eq.(\ref{Ltot})
have been computed in \cite{Gabrielli:2014oya}, as a function of the basic parameters of the model  described in Section~\ref{model}. 


In the $\bar{m}^2_L\simeq \bar{m}^2_R$ approximation of degenerate  messenger masses in the left and right-handed sectors,
 corresponding to the mixing angle $\theta=\pi/4$ (see \cite{Gabrielli:2013jka} for notations), the flavor-universal messenger mass matrix can be expressed in terms of  two parameters, the average messenger mass $\bar{m}^2=(m_L^2+m_R^2)/2$, and the mixing parameter  $\xi\equiv \Delta^2/\bar{m}^2$. We then define the universal mixing parameters $\xi_q$ and $\xi_l$, 
corresponding to the messenger mixing  parameters in the quark and lepton sectors,  respectively. Note that, 
in the effective theory approximation,  the Higgs and $Z$ masses  can be both set to zero in loop functions,  when terms  
${\cal O}(m_H^2/\bar{m}^2)$ are negligible.

Then, one finds\footnote{
Note that, due to the Bose statistics of the messenger fields, the relative sign with respect to the SM contribution in the 
$H\to gg$ amplitude is predicted to be negative. Analogously, there is a negative  relative sign with respect to the SM fermion contribution to the $H\to \gamma\gamma$ and $H\to \gamma Z$ amplitudes.}
\bea
C_{\gamma\bar{\gamma}}&=&\sqrt{\frac{\bar{\alpha}}{\alpha}}\sum_{i=q,l}\frac{R^i_1}{12}\frac{\xi_i^2}{1-\xi_i^2}\, ,\nonumber\\
C_{\bar{\gamma}\bar{\gamma}}&=&\frac{\bar{\alpha}}{\alpha}\sum_{i=q,l}\frac{R^i_2}{12}\frac{\xi_i^2}{1-\xi_i^2}\, ,\nonumber\\
C_{Z\bar{\gamma}}&=&\sqrt{\frac{\bar{\alpha}}{\alpha}}\sum_{i=q,l}R^i_{Z\gamma}\frac{R^i_1}{12}\frac{\xi_i^2}{1-\xi_i^2}\, ,\nonumber\\
C_{\gamma\gamma}&=&C^{\rm SM}_{\gamma\gamma}\left(1+\sum_{i=q,l}\frac{R^i_0\xi_i^2}{3F\left(1-\xi_i^2\right)}\right),\nonumber\\
C_{Z\gamma}&=&C^{\rm SM}_{\gamma\gamma}\left(1+ \sum_{i=q,l}R^i_{Z\gamma}\frac{R^i_0\xi_i^2}{3F\left(1-\xi_i^2\right)}\right),\nonumber\\
C_{gg}&=&C^{\rm SM}_{gg}\left(1+\frac{\xi_q^2}{3F_q\left(1-\xi_q^2\right)}\right),
\label{Ci}
\eea
where $C^{\rm SM}_{\gamma\gamma}=\frac{1}{8}F$, 
$C^{\rm SM}_{gg}=   \, \frac{1}{16}F_q\,$,
and the constants $R_{0,1,2}^{q,l}$ are given by
\bea
&& R^q_0=3N_c(e_{\U}^2+e_{\D}^2),~~~~~~~~~~~~~~~~~~~
R^l_0=3\,e^2_{\E}\, , 
\nonumber\\
&&R^q_1=N_c\sum_{i=1}^3\left(e_{\U} q_{\U_i}+e_{\D} q_{\D_i}\right), ~~~~~~~
R^l_1=e_{\E}\sum_{i=1}^3\left(q_{\EE_i}\right)\, , 
\nonumber\\
&&R^{\,q}_2=N_c\sum_{i=1}^3\left(q_{\U_i}^2+q_{\D_i}^2\right), ~~~~~~~~~~~~
R^{\,l}_2\, =\,\sum_{i=1}^3\left(q_{\EE_i}^2+q_{\nu_i}^2\right)\, ,
\label{Ri}
\eea
with $e_{\U}=2/3$, $e_{\D}=-1/3$, and $e_{\E}=-1$,  the  electric charges for up-, down-quarks, and charged leptons, respectively. $F$ and $F_q$ are the usual SM loop factor  given by 
\bea
F=F_W(\beta_W)+F_F\,,~~~~~\;\;\; F_F=\sum_f N_c Q^2_f F_f(\beta_f)\, ,
~~~~~~ F_q=\sum_f F_f(\beta_f)\, ,
\eea
with $N_c=1 (3)$ for leptons (quarks) respectively, 
$\beta_W=4m_W^2/m_H^2$, $\beta_f=4m_f^2/m_H^2$, and
\bea
F_W(x) &=& 2+3x+3x\left(2-x\right)f(x)\, , ~~~~~~
F_f(x) = -2x\left(1+(1-x)f(x)\right)\, ,
\eea
where $f(x)=\arcsin^2[\frac{1}{\sqrt{x}}]$, for $x\ge 1$, and 
$f(x)=-\frac{1}{4}\left(\log\left(\frac{1+\sqrt{1-x}}{1-\sqrt{1-x}}\right)-i\pi\right)^2$,  
for $x< 1$. Including  only the $W$ and 
top-quark loops in $F$,  we get 
$|F|\simeq 6.5$,  $|F_q|\simeq 1.25$ for  $m_H=125$ GeV. 
We will elaborate on the $R^{\,q,l}_{Z\gamma}$  constants in Eq.(\ref{Ci})  
(parametrizing the ratio of the messenger couplings to the $Z$ and $\gamma$) at the end of this section\footnote{We will  neglect  the $\bar{\alpha}$ running from the $\bar{m}$ scale to 
the characteristic low-energy scale entering  the dark-photon vertex in  $H\to \gamma\bar{\gamma}, \bar\gamma\bar{\gamma}, Z\bar{\gamma}$.}.

One can see that in the Higgs couplings in Eq.(\ref{Ci}), there is a clear non-decoupling effect,  since the $C_{i\,j}$ coefficients  do not vanish when  
$(\bar{m}^2, \Delta)\to \infty$, provided  the ratio $\Delta/\bar{m}^2$ is finite. 

The ratios $ r_{ij}$ ($i,j =\gamma , \bar {\gamma}$), defined  in Eq.~(\ref{rij}),  
entering  the  model-independent BR's parametrization in Eq.~(\ref{BRS}),
and $ r_{gg}$, defined in Eq.~(\ref{rgg}),
 are then given by
\bea
r_{\gamma\bar{\gamma}}&=&2\left(\sum_{i=l,q} X_i R^i_1\right)^2
\left(\frac{\bar{\alpha}}{\alpha}\right)\, ,~~~~~~
r_{\bar{\gamma}\bar{\gamma}}=\left(\sum_{i=l,q} X_i R_2^i\right)^2
\left(\frac{\bar{\alpha}}{\alpha}\right)^2\, , \\ 
r_{\gamma\gamma}&=&\left(\sum_{i=l,q} X_i R_0^i\right)^2\, , ~~~~~~~~~~~~~~~
r_{gg}=\frac{X_q^2F^2}{F_q^2}\, ,
\label{run}
\eea
where the extra factor 2 in $r_{\gamma\bar{\gamma}}$ comes from statistics and
\bea
X_{l(q)} \equiv \frac{\xi^2_{l(q)}}{3F(1-\xi_{l(q)}^2)}\, ,  
\label{rdue}
\eea
with $R^{q,l}_{0,1,2}$ defined in Eqs.~(\ref{Ri}).

The strength of the exotic contribution to 
$H\to \gamma\gamma$ is directly controlled by two  
 mixing parameters, $\xi_q$ and $\xi_l$. On the other hand, the $H\to gg$
depends only on $\xi_q$, and can be  constrained at the LHC by  measuring  Higgs production rates.

It is useful to 
connect the messenger-loop impact on the $Hgg$ vertex expressed by $r_{gg}$ [as defined in Eq.~(\ref{rgg})] with the usual $k_{g}$ anomalous coupling of the 
$Hgg$ interaction
which enters the relation  $C_{gg}=k_{g}\,C_{gg}^{\rm SM}$. By Eq.~(\ref{Ci}),
it is straightforward to see that $k_{g}\simeq1-\sqrt{r_{gg}}$.
Present data constraints $k_{g}$ at 68\% of C.L. to be in the ranges
$k_{g} = 1.00^{+0.23}_{-0.16}$ (ATLAS Collaboration \cite{ATLAS:h}),
and $k_{g} = 0.76^{+0.15}_{-0.13}$ (CMS Collaboration \cite{CMS:2014ega}).
In the following, we assume $r_{gg}$ to be in the range $0\lsim  r_{gg} \lsim 0.4$. 

\begin{figure}[t]
\begin{center}
\vskip -2.2cm
\hskip -1cm
\includegraphics[width=0.65\textwidth]{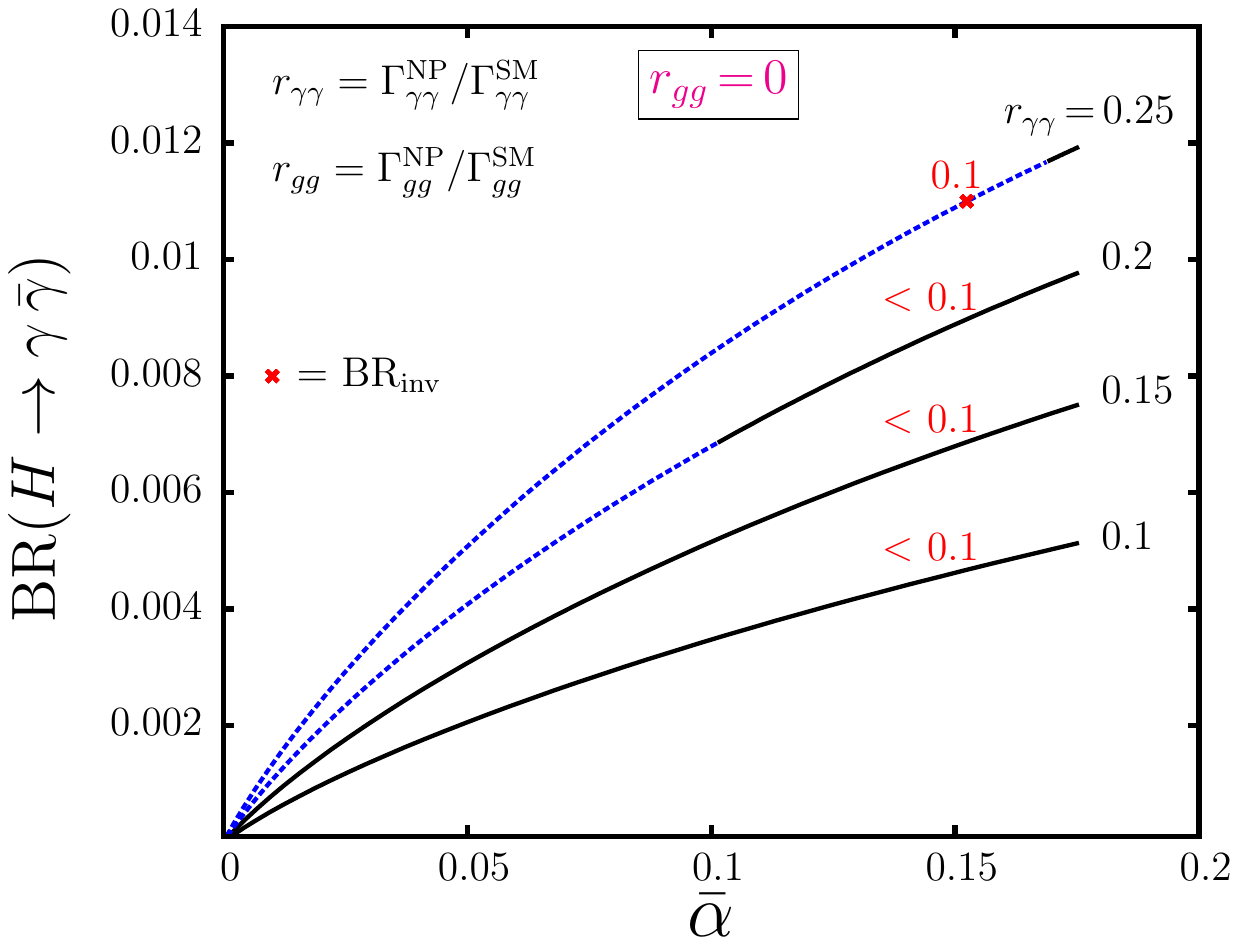}
\hskip -4.5cm
\includegraphics[width=0.65\textwidth]{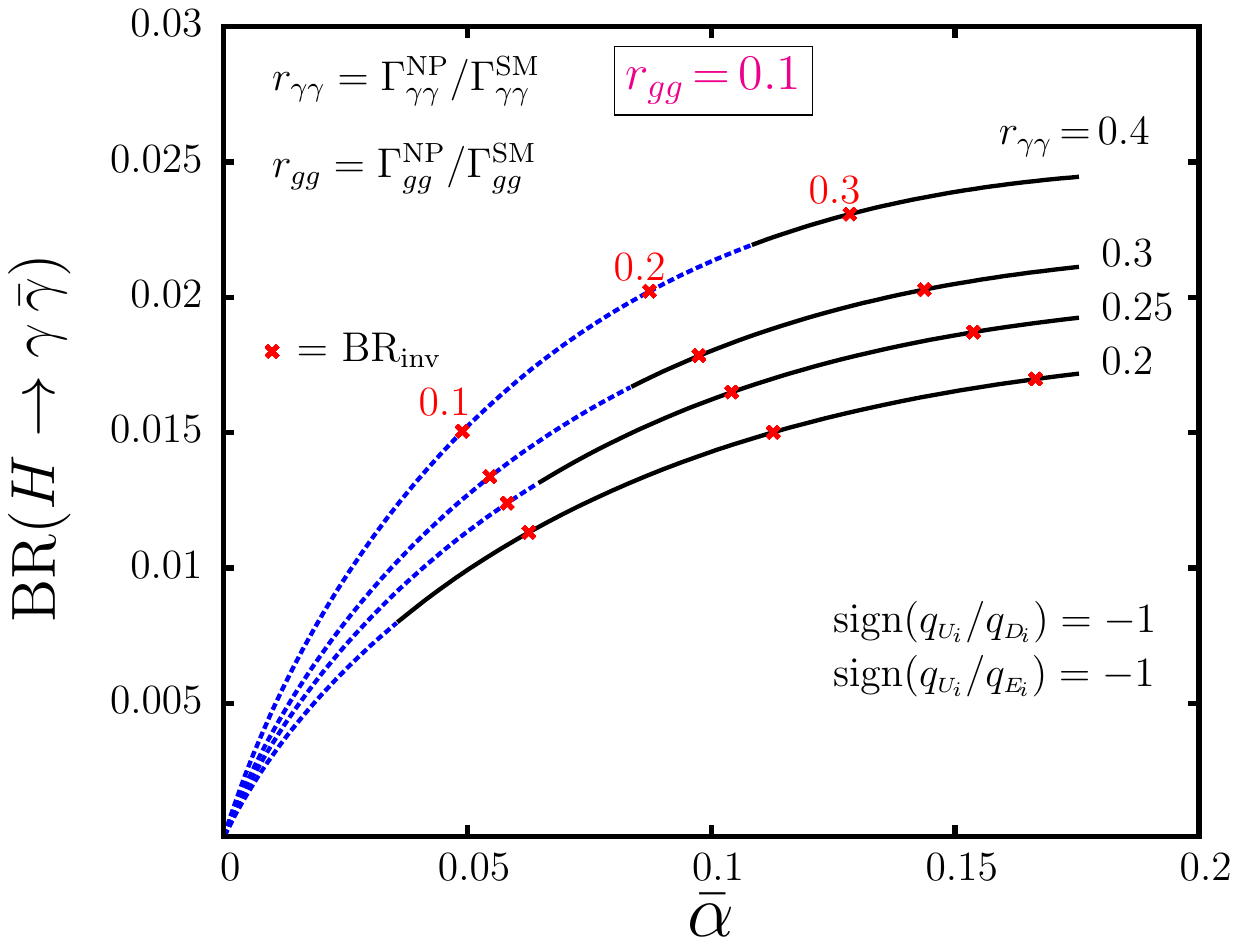} 
\vskip -9cm
\hskip -1cm
\includegraphics[width=0.65\textwidth]{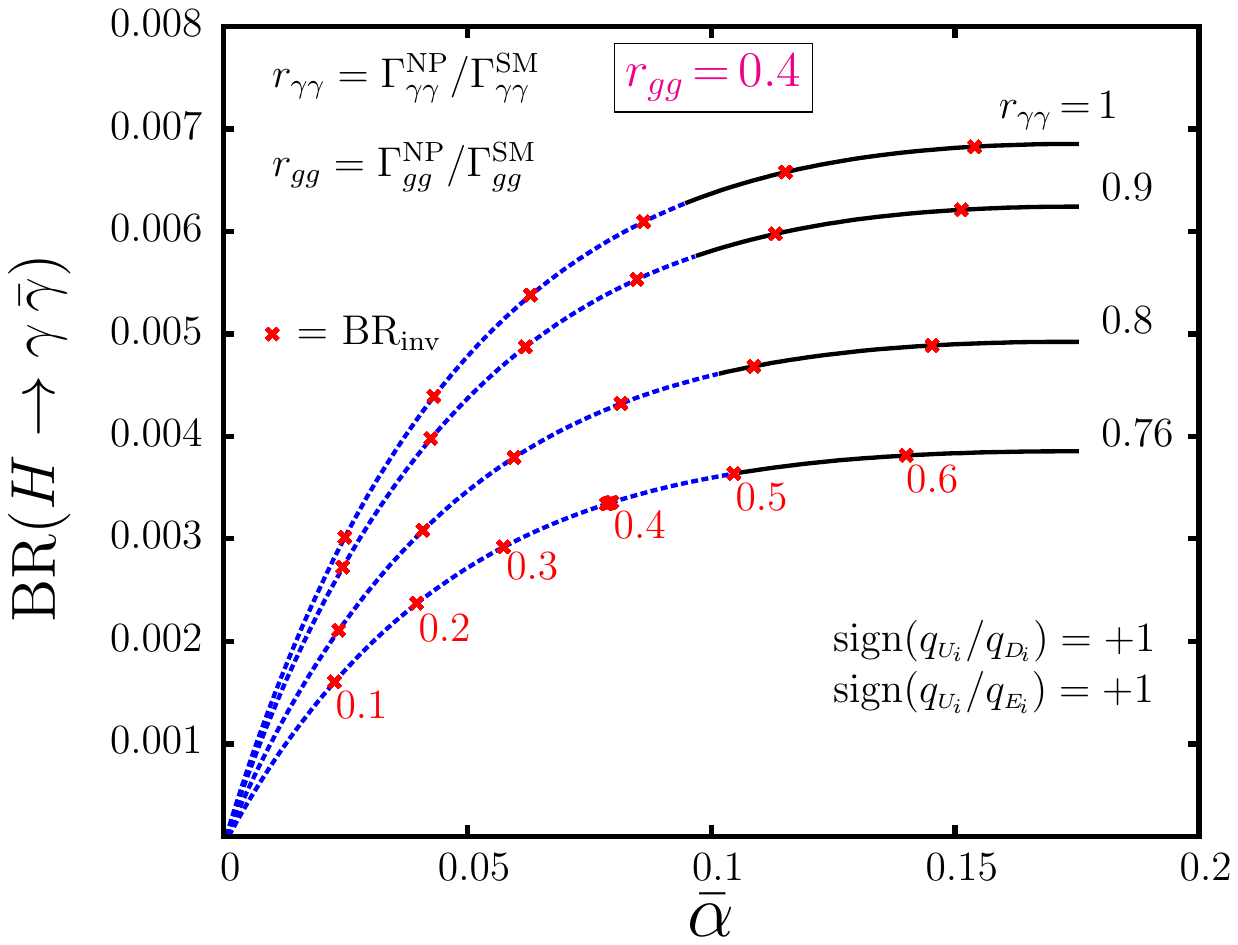}
\hskip -4.5cm
\includegraphics[width=0.65\textwidth]{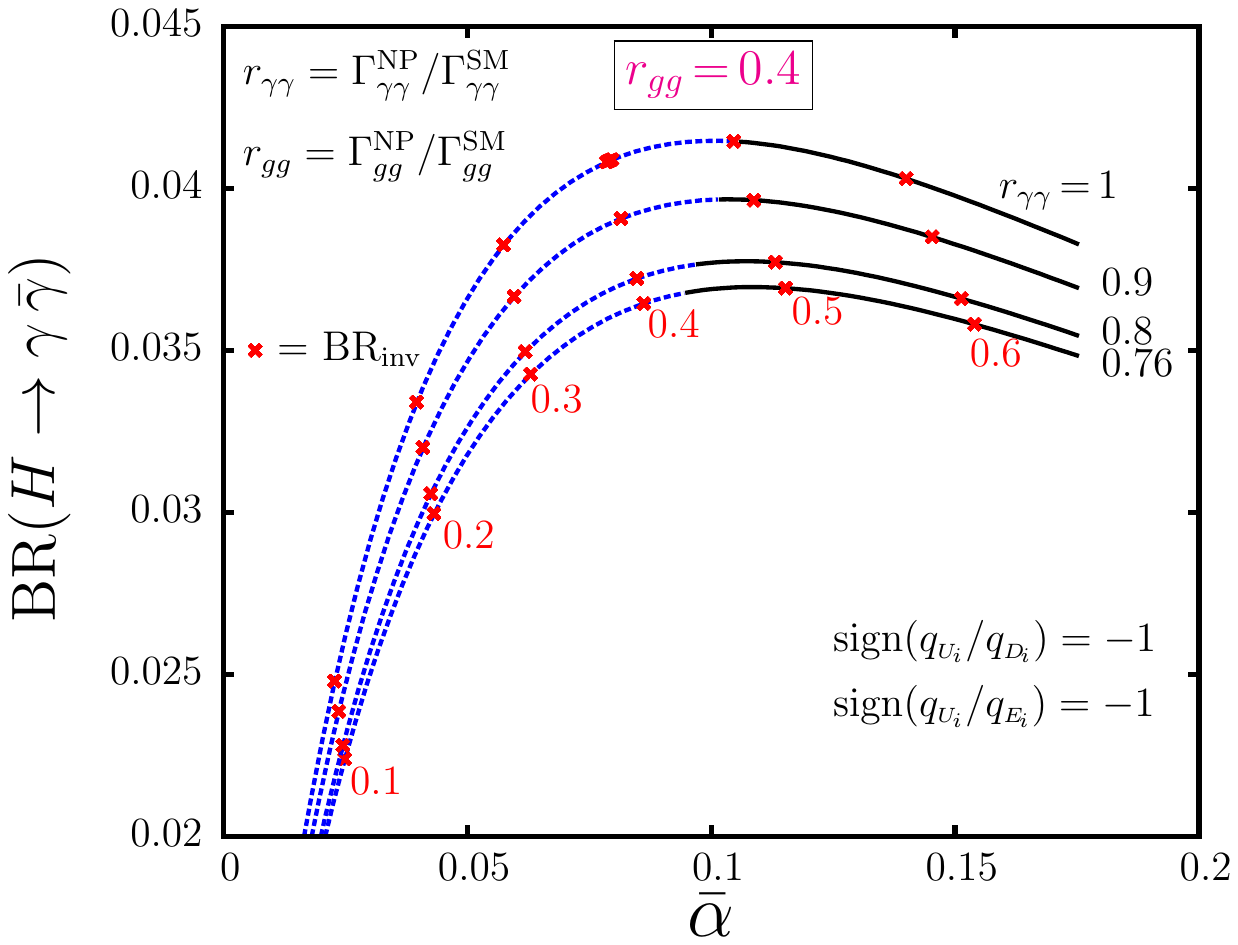} 
\vskip -7cm
\caption{Predictions for ${\rm BR}(H\to \gamma \bar{\gamma})$ as functions of
 the $U(1)_F$ fine structure constant 
 $\bar \alpha$,  
 for different values of ${\rm BR}_{\rm inv}$, $r_{\gamma\gamma}$  and  $r_{gg}$  in the full model; $r_{\gamma\gamma}$ satisfies the constraint
$r_{\gamma\gamma}\gsim 1.9 \,r_{gg}$ (see text).}
\label{fig11}
\end{center}
\vskip -0.3cm
\end{figure}

In Figure~\ref{fig11}, we show   BR($H\to \gamma\bar{\gamma}$)  
 as a function of the $U(1)_F$ fine structure constant  
evaluated at the average messenger mass, $\bar{\alpha}(\bar{m})$, for a few values of the ratios  $r_{gg}$ and $r_{\gamma\gamma}$. We assume the charge normalization  $q_{U_3}=1$ and $q_{L_3}=1$.
The dashed blue lines correspond to the condition
\bea
\frac{1}{2}{\rm BR^{\rm SM}}_{\gamma \gamma} <  {\rm BR}_{\gamma \gamma} <
2{\rm BR^{\rm SM}}_{\gamma \gamma}
\eea
with ${\rm BR}^{\rm SM}_{\gamma\gamma}=2.28\times 10^{-3}$.
The red dots  correspond to fixed BR$_{\bar \gamma \bar{\gamma}}\simeq {\rm BR}_{\rm inv}$ values, where  ${\rm BR}_{\rm inv}$ is the Higgs invisible-decay BR. Note that when colored messengers contribute to the 
$H\gamma \bar{\gamma}$ effective coupling,  BR$_{\gamma \bar{\gamma}}$ depends also on the sign of the $U(1)_F$ charges, which are  free parameters. 
Correspondingly,  
in Figure~\ref{fig11}, we  show cases in which the $U(1)_F$ charge sign 
in the quark sector gives either  destructive or  constructive interferences with the
electromagnetic charges. The maximum value of $\bar{\alpha}\simeq 0.18$ corresponds to $\bar{\alpha}(\Lambda)\simeq1$ (see Eq.(\ref{alpha-scale}) in the
Appendix for details).

In the upper-left plot in Figure~\ref{fig11}, 
we show   the $BR_{\gamma\bar{\gamma}}$ predictions for $r_{gg}=0$. The allowed  $BR_{\gamma\bar{\gamma}}$ values are at most 
about 
1\%,  and  one has  
$BR_{\gamma\bar \gamma}> BR^{\rm SM}_{\gamma\gamma}$ only for $r_{\gamma\gamma}\gsim 0.1$, corresponding to a quite large mixing parameter ($\xi_l>0.82$) in the leptonic messenger sector. The upper-right plot 
shows the case $r_{gg}\simeq 0.1$, assuming constructive $U(1)_F$ charge
interferences, corresponding to ${\rm sign}(q_{\U_i}/q_{\D_i})={\rm sign}(q_{\U_i}/q_{\E_i})=-1$. In this plot, the minimum $r_{\gamma\gamma}$ value  $\sim 0.2$  comes from  the 
constraint $r_{\gamma\gamma}\gsim 1.9\, r_{gg}$, arising from 
  Eqs.~(\ref{run})-(\ref{rdue}) as a result of 
the color and EW quantum 
numbers of the different amplitudes. 
$BR_{{\gamma}\bar\gamma}$ above 2\% are allowed in this case, corresponding to the range  $\bar{\alpha}\sim 0.10-0.18$, with $r_{\gamma\gamma} \sim 0.4$.

The lower-right plot matches the largest 
allowed contribution from colored messengers  
($r_{gg}\simeq 0.4$, with $r_{\gamma\gamma}\gsim 0.76$), which corresponds to a mixing parameter $\xi_q\simeq 0.88$, while the lower-left corresponds to $r_{gg}\simeq 0.2$ with $r_{\gamma\gamma}\gsim 0.38$.
The left and right plots correspond to 
the destructive and constructive effects of the $U(1)_F$ charges, that is 
 ${\rm sign}(q_{\U_i}/q_{\D_i})={\rm sign}(q_{\U_i}/q_{\E_i})=1$,
and  ${\rm sign}(q_{\U_i}/q_{\D_i})={\rm sign}(q_{\U_i}/q_{\E_i})=-1$, respectively.
 For large colored messenger contributions, 
the  actual $BR_{{\gamma}\bar\gamma}$ value dramatically depends on the charge signs.
In the case of constructive interferences,  $BR_{{\gamma}\bar\gamma}$  
reaches  the 3\%$-$4\% level, while in the destructive case it is 
always below 0.3\%.

Summing up,   large $BR_{{\gamma}\bar\gamma}$  values
up to 1\%$-$4\% are possible in this scenario. They
 correspond to the $\bar{\alpha}$  range expected from naturalness arguments
applied to $U(1)_F$ charges (see the Appendix for more details). 
This $BR_{{\gamma}\bar\gamma}$ range is equivalent to  
$C_{\gamma\bar{\gamma}}$  values  up to about 5 in the Lagrangian in Eq.~(\ref{Leff}), 
as shown in Figure~\ref{fig:BRversusC}. As we have seen, in Section 3, these 
$C_{\gamma\bar{\gamma}}$ values could be well inside the domain of sensitivity of the $e^+ e^- \to H \bar{\gamma}$ searches at future $e^+ e^-$ colliders 
(cf. Figures~\ref{fig:reach} - \ref{fig:significance}).

Before closing this section, we elaborate on the general connection between  the $HZ\bar{\gamma}$
and $H\gamma\bar{\gamma}$ couplings, induced at one loop.
This 
 will  depend on  the spin and 
the SM gauge-group representation of the  particles running in the loop.
On the other hand, 
here the $HZ\bar{\gamma}$ and $H\gamma\bar{\gamma}$ vertices are induced by scalar messenger fields in the $SU(2)_L\times SU(3)_c$ fundamental representation  (cf. Table~\ref{tab1}), which gives a definite prediction for the 
$HZ\bar{\gamma}$ and $H\gamma\bar{\gamma}$ coupling ratio, and for 
the $R^{\,q,l}_{Z\gamma}$  terms entering the $C_{Z\bar\gamma}$ 
and $C_{Z\gamma}$ effective couplings in Eq.(\ref{Ci}).

Then we can now motivate the $R_{Z\gamma}\simeq 0.79$ scenario that we considered  in the $e^+e^-\to H \bar{\gamma}$ study of Section 3.  $R_{Z\gamma}$ can be defined (in absence of scalar mixing between the doublet and singlet scalars circulating in  loops) by  the relation 
\bea
C_{ZX}=R_{Z\gamma}\,C_{\gamma X}\, ,
\eea
with $X= \gamma,\bar\gamma$, for the Higgs effective couplings 
in Eqs.~(\ref{Leff})-(\ref{LeffSM}).

The $Z$-boson coupling to a scalar particle $i$ is given by 
\begin{equation}
R^{\,i}_{Z\gamma}=\frac{(1-\frac{Y^i}{Q^i})-\sin^2\theta_W}{\sin\theta_W\cos\theta_W}
\end{equation}
times the photon coupling to the same scalar, where $Y^i$ and $Q^i$ are the 
hypercharge and  the electric charge of the scalar, and $\sin\theta_W$ is the Weinberg angle.

If the scalars in the loop are  SM particle {\it partners} (as happens in SUSY or in the model in~\cite{Gabrielli:2013jka}), they will share the quantum numbers of the left- and right-handed SM fermions. 

Then, for right-handed fermion partners ($Y^R=Q^R$), and 
for  left-handed electron partners
 ($Y^{\tilde{e}_L}=-\frac12$ and $Q^{\tilde{e}_L}=-1$),
  one has, respectively,  
\begin{equation}
R_{Z\gamma}^R=-\frac{\sin\theta_W}{\cos\theta_W}\simeq-0.55\, ,
~~~~~~~~~~
R_{Z\gamma}^{\tilde{e}_L}=\frac{\frac12-\sin^2\theta_W}{\sin\theta_W\cos\theta_W}\simeq0.64\,.
\end{equation}

The average $Z$-to-$\gamma$ coupling ratio for a pair of mass-degenerate right- and left-handed leptonic scalars is then 
\begin{equation}
R_{Z\gamma}^{\tilde \ell}=\frac{R_{Z\gamma}^{R}+R_{Z\gamma}^{\tilde{e}_L}}{2}\simeq0.045\,.
\end{equation}
For left-handed up- and down-type squarks (${Y^{\tilde{u}_L}}=\frac14{Q^{\tilde{u}_L}}$ and ${Y^{\tilde{d}_L}}=-\frac12{Q^{\tilde{d}_L}}$, respectively), one has instead
\begin{equation}
R_{Z\gamma}^{\tilde{u}_L}=\frac{\frac34-\sin^2\theta_W}{\sin\theta_W\cos\theta_W}\simeq 1.2,\qquad R_{Z\gamma}^{\tilde{d}_L}=\frac{\frac32-\sin^2\theta_W}{\sin\theta_W\cos\theta_W}\simeq 3.0\, .
\end{equation}
The average contribution from a mass-degenerate pair of right- and left-handed up and down squarks is then 
$R_{Z\gamma}^{\tilde{u}}=(R_{Z\gamma}^{R}+R_{Z\gamma}^{\tilde{u}_L})/{2}\simeq0.34$,
and $R_{Z\gamma}^{\tilde{d}}=(R_{Z\gamma}^{R}+R_{Z\gamma}^{\tilde{d}_L})/{2}\simeq1.23$, respectively. 
Assuming that also the up- and down-type scalars are mass degenerate,  the net result from a squark doublet is then
\begin{equation}
R_{Z\gamma}^{\,\tilde{q}}=\frac{R_{Z\gamma}^{\tilde{u}}+R_{Z\gamma}^{\tilde{d}}}{2}\simeq0.79\, .
\end{equation}
The same pattern  for  the $R_{Z\gamma}$ constants can be obtained in the model in \cite{Gabrielli:2013jka}, in the approximation of degenerate colored  messenger scalars. 

In Section 3, we include the case
$C_{Z\bar\gamma}=R^{\,\tilde{q}}_{Z\gamma}\,C_{\gamma\bar\gamma}=0.79\,C_{\gamma\bar\gamma}$ among the benchmarks for 
the  analysis of  the $e^+e^-\to H \bar{\gamma}$ potential, corresponding to negligible  leptonic contributions in the messenger loops.

\section{Conclusions}
Hidden sectors extending the SM theory can include an extra unbroken $U(1)$ gauge symmetry.
The corresponding gauge boson, a massless dark vector boson, can couple to the Higgs boson through renormalizable interactions involving scalar messengers, giving rise to
effective $H\gamma\bar{\gamma}$, $HZ\bar{\gamma}$, and $H\bar\gamma\bar{\gamma}$ couplings. Since a massless dark photon is not revealed by collider detectors, the latter can be probed at the LHC and future colliders via the search for exotic Higgs decays
into a photon or a $Z$ boson plus missing transverse energy, and the determination of the
 invisible Higgs decay width.
 
Another way to investigate the possible existence  of the $H\gamma\bar{\gamma}$ and  $HZ\bar{\gamma}$ couplings is the production at future $e^+e^-$ colliders of a Higgs boson
associated to a dark photon. The corresponding signature is very distinctive,
since there is no irreducible SM background where the Higgs decay products are accompanied by a massless invisible system.

After introducing an effective Lagrangian description of the new Higgs interactions, 
we studied the potential of the $e^+e^-\to H \bar{\gamma}$ production for probing 
the corresponding $H\gamma\bar{\gamma}$ and  $HZ\bar{\gamma}$ couplings, 
$C_{\gamma\bar{\gamma}}$ and $C_{Z\bar{\gamma}}$.
A most useful strategy for enhancing the $S/B$ ratio turns out to be a selection on the small 
values of the invisible-system
invariant mass.
We found that, at $\sqrt s = 240$ GeV, with the integrated luminosity foreseen at the FCC-ee ($\sim10$ ab$^{-1}$),
one can exclude at 95\% C.L. the ranges $C_{\gamma\bar{\gamma}}>1.9$
(for $C_{Z\bar{\gamma}}=0$), $C_{Z\bar{\gamma}}>2.7 $ (for $C_{\gamma\bar{\gamma}}=0$), and $C_{\gamma\bar{\gamma}}>1.6$ (for
$C_{Z\bar{\gamma}}=0.79\;C_{\gamma\bar{\gamma}})$.
The interval $C_{\gamma\bar{\gamma}}>1.9$ corresponds to   BR$(H\to
\gamma\bar\gamma)> 3$    BR$_{SM\!}(H\to \gamma {\gamma})$, while, excluding $C_{Z\bar{\gamma}}>2.7 $,  one excludes  BR$(H\to Z\bar\gamma)> 9$  BR$_{SM\!}(H\to Z {\gamma}$).
The corresponding BR bounds for the Higgs decay into a dark photon at the ILC with $\sqrt s = 250$ GeV, and $\sim250$ fb$^{-1}$
of integrated luminosity, are about an order of magnitude looser than the FCC-ee ones.

Of course, in order to fully assess the $e^+e^-$ collision potential, one would need
an estimate of the corresponding LHC sensitivity to the $H\gamma\bar{\gamma}$ and  $HZ\bar{\gamma}$ couplings. This would in particular allow to figure out whether, in case of negative findings at  the forthcoming LHC runs, 
there will indeed be some unexplored parameter space  left that can be covered by
searches in $e^+e^-$ collisions. Studies to evaluate  the LHC  sensitivity to the
$H\to \gamma\bar\gamma$ channel at Run 1 have just been started.
 A  parton-level analysis leads to a bound on BR$(H\to
\gamma\bar\gamma)$ of about half a percent at 95\% C.L. from the 8$-$TeV data set \cite{Gabrielli:2014oya}, corresponding to the  exclusion of the  range $C_{\gamma\bar{\gamma}}>1.6$ (cf. Figure~\ref{fig:BRversusC}).
On the other hand, since most of the background to the $H\to \gamma\bar\gamma$
signal in $pp$ collisions comes from 
jet- and photon-mismeasurement effects,  a thorough detector simulation would be needed to make a robust sensitivity statement on the present data set at the LHC. Studies at 13$-$14 TeV
are expected to increase the sensitivity on a purely signal-statistics basis, 
but will be affected by harsher experimental conditions, that make the
reconstruction of relatively small missing transverse energies  quite critical. 
The extrapolation of  present 8$-$TeV results to larger $pp$ c.m.  energies will then require an even more careful analysis. Similar considerations apply to the LHC searches for 
 a $H\to Z\bar\gamma$ signal, where the expected missing transverse energy
 is even  lower than in the $H\to \gamma\bar\gamma$ case.

Predictions for the BR($H\to \gamma \bar{\gamma}$) in the framework of the Flavor model  proposed in \cite{Gabrielli:2013jka} have also  been presented. Due to  non-decoupling effects,  BR($H\to \gamma \bar{\gamma}$) turns out to be directly proportional to the mixing parameters $\xi_{q,l}$ in the scalar messenger sector, and  to  the $U(1)_F$ coupling constant $\bar{\alpha}$, times some SM couplings. Remarkably, quite 
high $\xi_{q,l}$ mixings are  required in order to  generate the Yukawa couplings 
radiatively,  and avoid  large fine 
tuning in the Higgs sector \cite{Gabrielli:2013jka}. At the same time, 
 large (but still in the perturbative range) $\bar{\alpha}$ couplings are expected, in order to avoid fine-tuning among the dark fermion charges.
As a consequence, in the scenario
 \cite{Gabrielli:2013jka}, BR($H\to \gamma \bar{\gamma}$) can be naturally large, and in the ballpark of sensitivity of the LHC and future 
 colliders.

In conclusion, although realistic LHC sensitivity studies for the Higgs couplings to dark photons are still to come, one expects quite a lot of complementarity of the LHC and future $e^+e^-$ capabilities to probe  new exotic Higgs-boson interactions with dark photons, thanks to the extremely clean $e^+e^-$ experimental conditions.

\vskip 0.9cm

\mysection{Acknowledgments}
E.G. would like to thank the PH-TH division of CERN for its kind 
hospitality during the preparation of this work.
This work was supported by the ESF grants  MTT59, MTT60,
by the recurrent financing project SF0690030s09, and by the European Union through the European Regional Development Fund.

\newpage
\section*{Appendix}
One of the most interesting property of the Flavor model in \cite{Gabrielli:2013jka} is that the value of the $U(1)_F$ coupling strength in the dark sector, $\bar{\alpha}$, can be  connected to the charge splitting among two dark-fermion generations in the lepton or quark sector. In the following, we present such a prediction.
 
A crucial assumption in the model is the one of minimal flavor violation,  
implying that the only source of flavor violation comes from the $U(1)_F$ charges. 
We first define $\bar{\alpha}(\Lambda)$ by normalizing to 1 the largest $U(1)_F$ charge,
which is the one associated to the dark-fermion partner of the top quark. Then, according to Eq.(\ref{Yeff}), we get
\bea
1/\bar{\alpha}(\Lambda) &\simeq &\frac{3}{2\pi} \frac{q_{Q_i}^2}{1-q_{\Q_i}^2}
\log\frac{m_t}{m_i}
\label{sumrules1}
\eea
where $m_t$ is the top-quark mass,  $m_i$ stands for another generic quark mass, and $q_{\Q_i}$ is the  $U(1)_F$ charge of the  corresponding dark-fermion partner 
$Q_i$\footnote{We have used the property that the loop  function $Y_0(x_i)$, as defined in Eq.(\ref{Yuk}), has a weak dependence on the dark-fermion mass $M_{Q_i}$, and can be approximated  to a constant.}.
Then, by fixing  the ratio of two $q_{\Q_i}$ charges in Eq.(\ref{sumrules1}), one can predict $\bar{\alpha}(\Lambda)$, as well as all the remaining $U(1)_F$ charges.
For instance, by requiring a charge splitting  of $10\%$ among the  $U(1)_F$ charges of  the dark-fermion partners of the third quark generation, namely $q_{\U_3}\simeq 1$ and $q_{\D_3}\simeq 0.9$, we get 
$\bar{\alpha}(\Lambda) \simeq  0.13$, and also $q_{\U_2}\simeq 0.87$, $q_{\D_2}\simeq 0.82$, and $q_{\U_1}\simeq 0.78$, $q_{\D_1}\simeq 0.76$,
where for the different quark masses we assumed the central values in \cite{pdg}.  

On the other hand, in order to obtain a weakly coupled 
$U(1)_F$ theory, with $\bar{\alpha}(\Lambda) \ll 10^{-2}$, a charge splitting 
$\ll 1\%$ is required, leading to an unnatural  fine tuning among the 
$U(1)_F$ charges. 
Remarkably, the same conclusion holds when extracting $\bar{\alpha}(\Lambda)$ from   the Eq.~(\ref{sumrules1}) applied to the purely EW dark sector. In this case, assuming  $q_{\EE_3}=1$ and $q_{\EE_2}=0.9$ for the dark fermions associated to the  $\tau$  and $\mu$ leptons, respectively, one obtains $\bar{\alpha}=0.17$, which is
of the same order of the coupling strength for dark fermions in the quark sector.
In order to avoid an unnatural  fine-tuning among the $U(1)_F$ charges, Eq.(\ref{sumrules1}) suggests a large, but maybe still perturbative, $U(1)_F$ coupling in the dark sector.

\begin{figure}[t]
\begin{center}
\vskip -2.6cm
\includegraphics[width=0.8\textwidth]{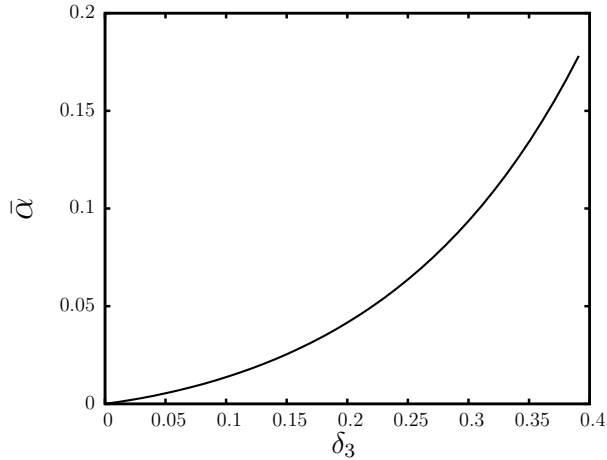}
\vskip -8.5cm
\caption{Predictions for 
 $\bar \alpha$ at the low-energy messenger scale $\bar{m}$,
 as a function of the  $U(1)_F$ charge splitting in the third generation, defined by $q_{Q_{\D_3}}=q_{Q_{\U_3}}(1-\delta_3)$.}
\label{fig0}
\end{center}
\vskip -0.3cm
\end{figure}

Actually, in \cite{Gabrielli:2013jka} the scale $\Lambda$ can be 
many orders of magnitude above the messenger mass scale, and it is useful 
to have the $\bar{\alpha}$ value   at low energy (for instance, at the average messenger mass $\bar{m}$),  as a function of the charge splitting of a pair of different dark fermions. Indeed, it is the low-energy $\bar{\alpha}$   that  enters BR($H\to \gamma \bar{\gamma}$) and  BR($H\to \bar{\gamma} \bar{\gamma}$).
We then need first to connect  $\bar{\alpha}(\Lambda)$
 to  $\bar{\alpha}(\bar{m})$ by solving the appropriate
renormalization group equations for the $U(1)_F$ $\beta$-function.
Due to the large energy gap between  $\Lambda$ and $\bar{m}$, 
  all dark-fermion  and messenger masses can  be approximated to a common low-energy scale around $ \bar{m}$, neglecting the running between different mass thresholds at low energy.
Then,  including the dark-fermion and
messenger contributions to the one-loop beta-function, as well as the  dark-fermion mass definition  in Eq.(\ref{mgap2}), one can  remove  the explicit dependence on the $\Lambda$ scale, by the expression  
\bea
\bar{\alpha}(\Lambda)=\bar{\alpha}(\bar{m})\left(  
1+R_2^{\,q}\left(\frac{4+3 N_c^2}{9N_c}\right)
+R_2^{\,l}\frac{7}{9}\right)\, ,
\label{alpha-scale}
\eea
where $R^{\,q,l}_2$ are defined in Eq.(\ref{Ri}). 
The peculiar solution in 
Eq.(\ref{alpha-scale}) arises from reabsorbing the usual  
$\log(\Lambda/M_{Q_i})$ term into the dark-fermion mass definition  in Eq.(\ref{mgap2}).

In Figure~\ref{fig0}, the $\bar{\alpha}$ behavior is shown   
as a function of the  $U(1)_F$ charge splitting $\delta_3$ 
associated to the
third generation of quarks,  with $q_{Q_{\D_3}}=q_{Q_{\U_3}}(1-\delta_3)$.  We assume the charge normalization  $q_{U_3}=1$, and  also  $q_{L_3}=1$ for 
the $U(1)_F$ charges of the dark fermions and messengers of the 
third-generation  leptonic sector.
The end point at $\delta_3=0.39$ corresponds to 
$\bar{\alpha}(\Lambda)=1$. 

By requiring a {\it natural} charge splitting that is  not  smaller than 20\%, $\bar{\alpha}$ 
turns out to be quite strong,  $0.04 < \bar{\alpha} < 0.18$, but still in the perturbative regime. Then, one obtains quite naturally large values for the $H\to \gamma \bar{\gamma}\;$ branching ratio.


\begin{thebibliography}{99}
\bibitem{CMS:2014ega}
  CMS Collaboration,
  ``Precise determination of the mass of the Higgs boson
and studies of the compatibility of its couplings with the standard model,''
  CMS-PAS-HIG-14-009.
\end{thebibliography}
\end{document}